\documentclass[11pt]{article}
\usepackage{moriond,epsfig}
\usepackage{graphicx}
\pdfoutput=1

\def\be{\begin{equation}}
\def\ee{\end{equation}}
\def\bea{\begin{eqnarray}}
\def\eea{\end{eqnarray}}

\begin{document}
\vspace*{4cm}
\title{MORIOND 2012, QCD AND HIGH ENERGY INTERACTIONS -EXPERIMENTAL SUMMARY-}

\author{ G. Dissertori }

\address{Institute for Particle Physics, ETH Zurich,\\
Schafmattstr. 20, 8093 Zurich, Switzerland}

\maketitle\abstracts{
The Moriond conference on QCD and High Energy Interactions has
been a most exciting and interesting event once more. This year's
edition has been characterized by a very large amount of
new results from the LHC experiments. However, also
the experiments at other present or past accelerators are still giving extremely important input
to our quest for the understanding of nature at shortest distance
scales. In this review I will attempt to summarize the main 
experimental highlights of this conference.
}

%%%%
%%%%%%%%%%%%%%%%%%%%%%%%%%%% the Intro
%%%%
\section{Introduction}
 \label{sec:intro}
The focus of the Moriond ``QCD and High Energy Interactions''
conference series is on theoretical and experimental advances in our
understanding of the Standard Model (SM), by studying processes up to
the highest achievable energy scales, together with searching for
physics beyond the SM. Particular attention is given to the many
aspects of Quantum Chromodynamics (QCD), the theory of strong
interactions. Phenomena ranging from strongly interacting matter
at high energy densities, to scattering processes at highest energies
and thus shortest distance scales, are being probed at unprecedented
experimental precision and thus provide extremely valuable input to
the theoretical community. Furthermore, the study of hadrons
containing heavy quarks provides deep insights into phenomena related
to the CP-violating sector of the SM. 

This year's conference \cite{weblink} has been characterized by a
most impressive amount of results presented by the LHC experiments. 
Most of these new measurements are based on the statistics collected
during the 2011 LHC run. Also the TEVATRON experiments continue
to be an important player in the field, with new results appearing,
which are based on the full Run II dataset of about 10
fb$^{-1}$. These being hadron (as well as heavy ion) colliders,
obviously a deep understanding of QCD at many energy scales is
necessary. The rich spectrum of LHC and TEVATRON physics is
complemented by new results from past and present accelerators.
Here an overview of the most recent developments will be given, 
which by the nature of the vast richness of the field cannot be
complete. The author thus apologizes for any important omission.

The structure of this summary is the following. Our main tools, namely accelerators and detectors, are listed in
Section \ref{sec:tools}. 
Section \ref{sec:HI} is dedicated to Heavy Ion Physics, and in Section \ref{sec:protons}  we discuss new results
on the proton structure and inelastic proton scattering. 
Heavy Flavour Physics is addressed in Section \ref{sec:HF}, followed by a 
summary of recent tests of perturbative QCD in Section  \ref{sec:QCD}.
Top Physics is discussed in Section \ref{sec:TOP}. Finally, searches for Physics beyond
the SM and for the Higgs boson are summarized in Sections \ref{sec:NewPhen}
and \ref{sec:Higgs}, respectively. 
Regarding the status of theoretical advances in the field, we refer to a dedicated
review \cite{Soper}  in these proceedings.

%%%%
%%%%%%%%%%%%%%%%%%%%%%%%%%%% our tools
%%%%
\section{Our Tools}
 \label{sec:tools}

None of the results presented below would have
been possible without the excellent performance of our tools, namely
the accelerators and detectors. While at such conferences typically
only the final results of long analysis chains are shown, it is easy
to forget and praise all the immense work and ingenuity, which has
gone into the design, construction, commissioning and operation of
the various accelerators and the corresponding experiments.

Currently, the world's spotlights are
on the LHC and its experiments. As presented by Lamont \cite{Lamont}, 
the LHC machine physicists and
engineers had many special events to celebrate during the year 2011, 
because of several important milestones and even world records
achieved, mostly in terms of beam intensities, instantaneous and
integrated luminosities, both for the p-p and the heavy ion (HI)
running. Overall, during the proton run the LHC has delivered about 12.5 fb$^{-1}$ to its
experiments, with the largest fraction (about 5.5 fb$^{-1}$ each) to 
the two general purpose detectors ATLAS and CMS, and smaller
amounts, because of luminosity leveling, to LHCb (1.2 fb$^{-1}$)
and ALICE  (0.005 fb$^{-1}$). This became possible thanks to an increase by
a factor of 20 in the p-p peak luminosity, compared to the 2010
run. A similar factor of 20 increase has been achieved for the
integrated luminosity of the HI run in 2011, compared to 2010, 
with 150 $\mu$b$^{-1}$ of Pb-Pb collisions delivered.

For the 2012 run it has been decided to increase the centre-of-mass
energy to 8 TeV and to stay with 50 ns bunch spacing. Very tight collimator
settings should allow for regularly running at  a  $\beta^*$ value of 0.6 m
at the ATLAS and CMS interaction points. Combined with bunch
intensities above the design value of $1.1\times10^{11}$ it is planned
to attain peak luminosities of close to $7\times10^{33}$cm$^{-2}$s$^{-1}$.
Taking the planned $\sim126$ days of p-p running and the expected
beam parameters above,
integrated luminosities anywhere between 12 and 19 fb$^{-1}$ can be
expected for ATLAS and CMS at the end of 2012. For LHCb and ALICE a
larger $\beta^*$ value of 3 m is foreseen. Finally, at the end of the
2012 p-p run, currently some 22-24 days of p-Pb running are scheduled.
It is worth noting that at the time of writing this review the LHC has
already gone very close to the 2012 milestones in terms of peak
luminosity, a fantastic achievement of the very short and smooth
intensity ramp-up for this new 8 TeV run.

However, results shown at this conference were not only based on LHC
data. Up to its end of operations in September 2011, 
the TEVATRON has delivered the impressive amount of 12 fb$^{-1}$
of p-$\bar\mathrm{p}$ collisions, with ~10 fb$^{-1}$ recorded and 
already analyzed to a large extent by the CDF and D\O\ experiments.
New results are still arriving from the HERA experiments, as well as
from the B-factories and DAPHNE. The BESIII e$^+$e$^-$ ring by now has delivered
the world's largest samples of $J/\psi$ and $\psi'$ mesons. The RHIC 
experiments have presented new results on HI collisions, and new
measurements were presented based on data from the fixed-target SPS experiment
COMPASS and even from the former JADE detector at PETRA.

%%%%
%%%%%%%%%%%%%%%%%%%%%%%%%%%% Heavy Ions
%%%%
\section{Heavy Ion Physics}
 \label{sec:HI}

One of the main purposes of HI physics is to study the
properties of a quark-gluon plasma (QGP), supposed to have existed
during the initial stages of the Universe and to be created, under
laboratory conditions, in HI collisions. 
More generally,
the aim of HI experiments is to gain a better understanding
of the many phases, which the strongly interacting matter
undergoes from its initial creation in the HI collision to the
freeze-out of colour-neutral hadrons. Having experiments
at different centre-of-mass energies and with different colliding
particles, such as at RHIC and LHC, gives additional handles for
probing various areas in the phase diagram. Also, more and
more (hard) probes are being studied, 
in particular at the LHC, where unprecedented measurements
based on jets, photons, Z and W bosons have been made.
Combining the information obtained by different probes and
observables, the goal is to shed light on properties such as
the hydrodynamic behaviour of the hot and dense medium,
jet quenching, quarkonia suppression and the medium's 
``transparency'' to colored partons or colour-neutral
electromagnetically or weakly interacting particles.

\begin{figure}[htb]
\centering
\begin{tabular}{cc}
\includegraphics[width=7.5cm]{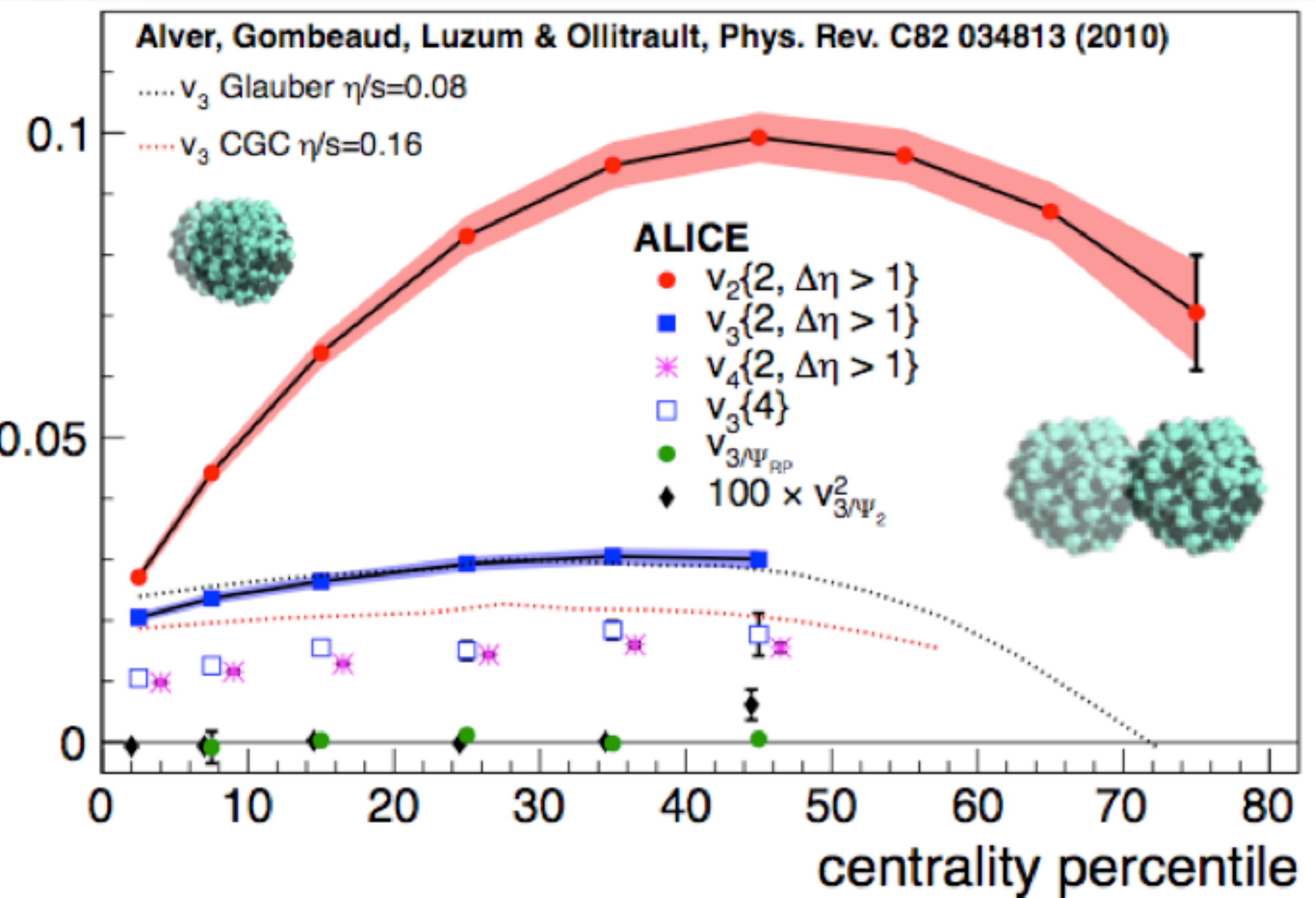} &
\includegraphics[width=6cm]{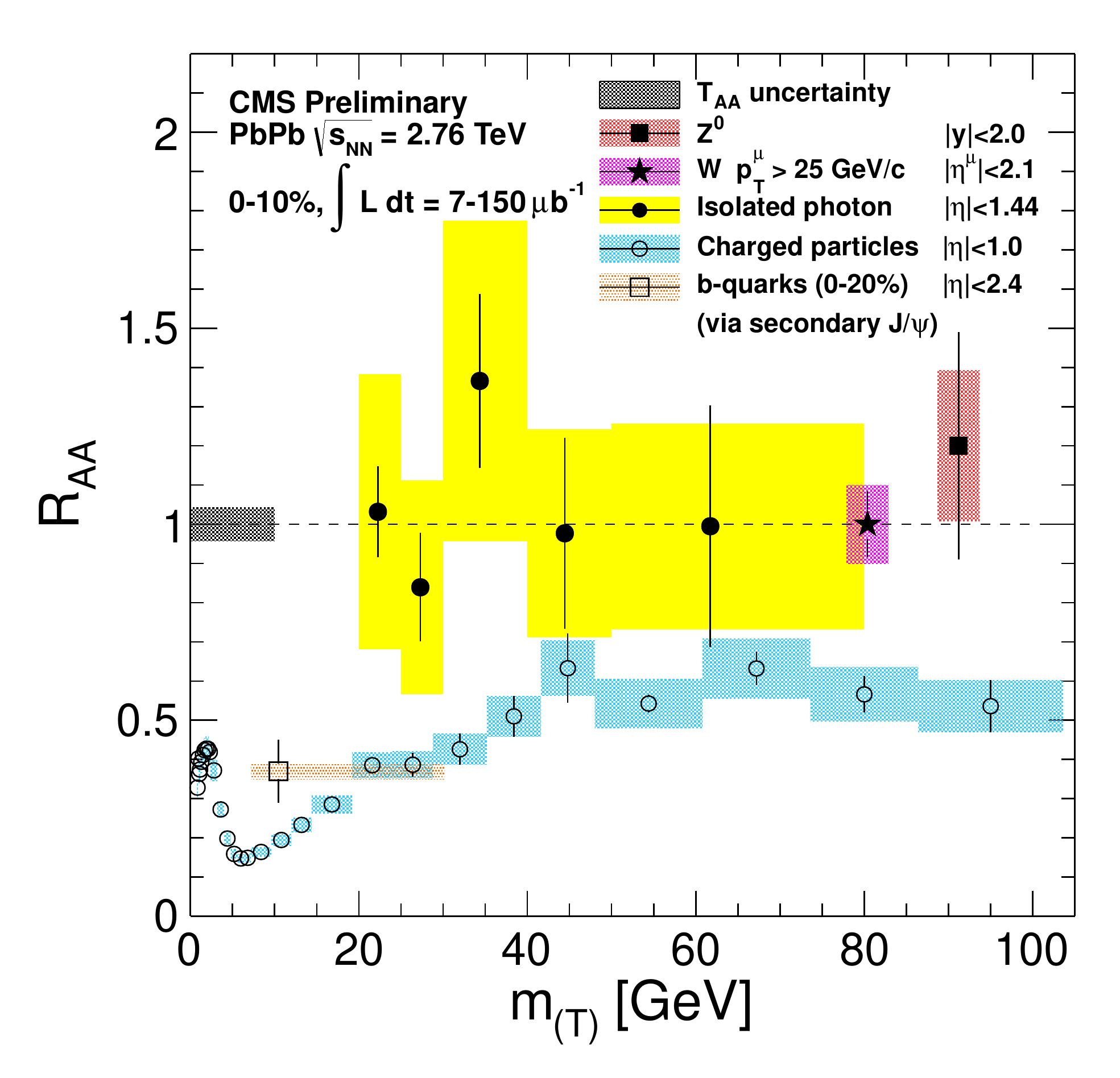} 
\end{tabular}
\caption{Left: Higher harmonics in flow measurements by ALICE,
              as a function of the centrality percentile;
             Right: $R_{AA}$ measurements by CMS for charged
              particles, b-quarks, photons, W and Z bosons.}
\label{fig:HI}
\end{figure}

An important class of measurements is given by the study
of hydrodynamic {\em{flow}}. Starting from the simple fact that
the initial state should exhibit a spatial anisotropy, due to 
non-central HI collisions, the pressure-driven expansion is
expected to lead to an anisotropy in the distribution of
 final state momenta or correlations thereof. A standard approach
to quantifying these anisotropies is by expanding such distributions
in terms of a Fourier series in the angle w.r.t. the reaction plane. Whereas the
2nd order coefficient ($v_2$) captures the elliptic nature of the
flow, and is well established since years, precise measurements
of non-vanishing higher-order moments have become available 
only recently (see, e.g.\ Fig.\ \ref{fig:HI}, left), giving important insight \cite{Snellings}.
Indeed, the fact that such higher harmonics are non-zero gives
strong support to the hydrodynamic picture of the QGP as a perfect
liquid, whereas a finite viscosity of the medium would smear
out any anisotropies due to fluctuations in the initial state and
only lead to a finite $v_2$ term. Quite a
number of new results from flow studies have been shown
by ATLAS, CMS, PHENIX, STAR and ALICE, with first measurements
up to very high transverse momenta, flow distributions for
identified hadrons and anti-particles, di-hadron correlations 
and first attempts to measure $v_2$ with di-leptons, photons and D
mesons. We refer to the dedicated HI contributions in these
proceedings for further details.

As mentioned above, hard probes have become quite a unique tool
for the LHC experiments. Here the main observable is the
so-called nuclear modification factor $R_{AA}$, a ratio of properly scaled A-A
and p-p cross sections in order to account for the number of nucleon
interactions in a single HI (A-A) collision. If a particular particle
(probe) is not affected by the presence of the dense and hot medium,
this ratio is expected to be unity. However, a strong suppression
below 1  is expected and observed for hadrons, due to the influence
of the medium on the colored constituents and thus on the hadron
formation. Earlier RHIC measurements have been extended by the LHC experiments to $p_T$
values up to 50 GeV  \cite{Knichel} or higher \cite{Yilmaz}, where the initial strong rise of the
modification factor between 10 and 20 GeV is observed to flatten
out above $\sim40$ GeV. This behaviour is captured by a large number of models, however,
the large spread of model predictions indicates the need of still a
better understanding to be obtained. First measurements of 
$R_{AA}$ for photons, Z and W bosons have been presented by ATLAS
\cite{Grabowska} and CMS \cite{Yilmaz}, cf.\ Fig.\ \ref{fig:HI}, right. As expected,
$R_{AA}$ values consistent with 1 within the still relatively large
uncertainties are found. Interestingly, ATLAS has also measured the
ratio for W and Z production, $R_{W/Z}=10.5\pm2.3$ and found it to be consistent with the SM
expectation, within the large uncertainty. Such types of measurements,
with better statistics, should allow to obtain information on the
parton distributions functions (pdfs) and their modifications in HI
collisions. Concerning the observation of quarkonia suppression, it
has been concluded that still a number of questions have to be
clarified before firm statements can be made, in particular those
related to the understanding of the initial conditions. Here dedicated
RHIC studies at forward rapidity will give important input \cite{Perdekamp}.
Finally, the clear observation of jet quenching by ATLAS and CMS,
reported after the 2010 run, has been confirmed and studied in more
detail thanks to the large 2011 statistics. For example, CMS has found
that an enhanced imbalance exists at all jet transverse momenta
\cite{Yilmaz}, while at the same time no angular decorrelation and no
significant modification of the jet fragmentation function is observed.

%%%%
%%%%%%%%%%%%%%%%%%%%%%%%%%%% Proton structure
%%%%
\section{Proton Structure}
 \label{sec:protons}

\begin{figure}[htb]
\centering
\begin{tabular}{lr}
\includegraphics[width=6cm]{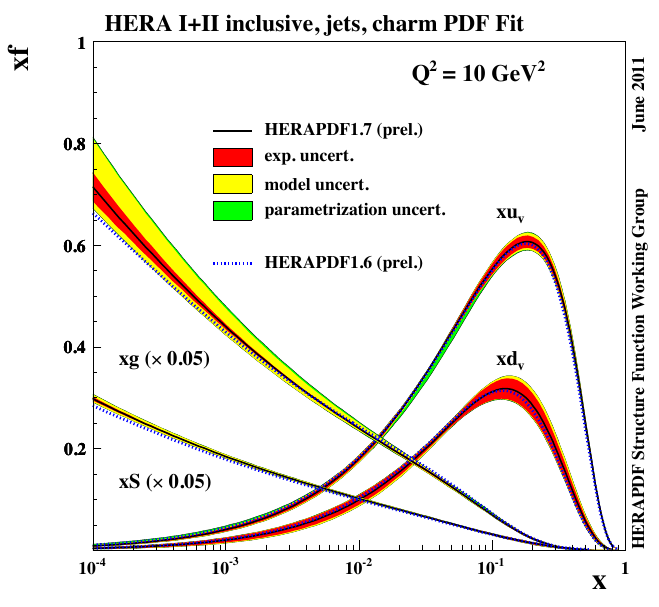} &
\includegraphics[width=7.5cm]{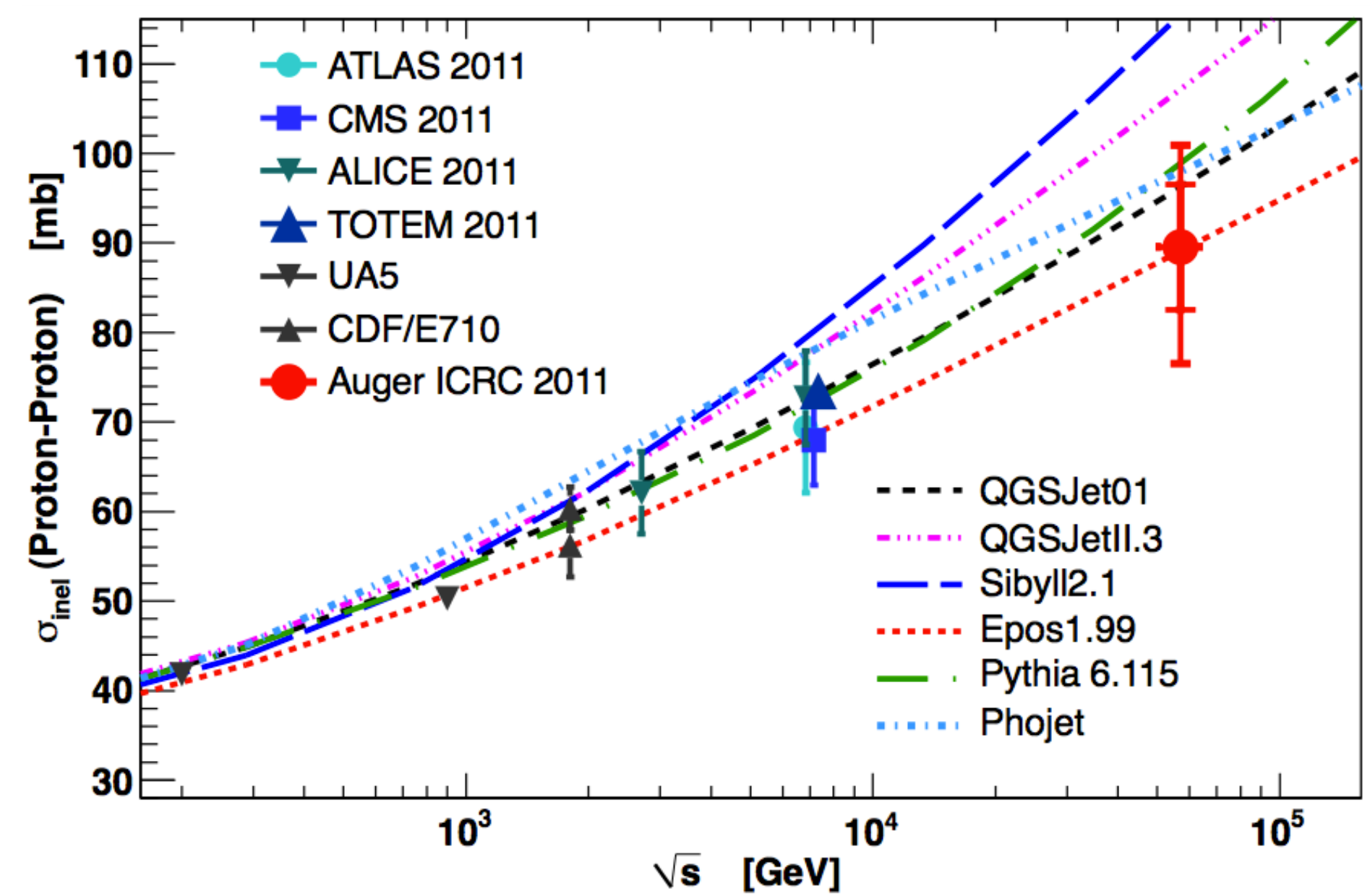} 
\end{tabular}
\caption{Left: Latest HERA pdf set, based on inclusive and charm structure functions and jet data;
               Right: Comparison of inelastic proton cross section data, from collider experiments and Auger, with
                various models.}
\label{fig:proton}
\end{figure}

The HERA and SPS experiments have presented some of their legacy measurements, in terms of
unpolarized and polarized (spin-) structure functions and their interpretations.
For example, a new and precise determination of HERA pdfs has been shown \cite{Lontkovskyi}
(HERAPDF1.7, cf.\ Fig.\ \ref{fig:proton} left). It is based on
combined inclusive HERA I + HERA II neutral and charged current  data,
HERA jet data, which reduce the strong correlation between the strong coupling
$\alpha_s$ and the gluon pdf, and a combined $F_2^{c\bar c}$ measurement, which
gives sensitivity to the gluon and charm content of the proton. It is worth highlighting that
predictions, based on such proton pdfs extracted from $e^{\pm} p$ data alone,
actually provide a good description of the LHC data. 

Extensive account of the wealth
of pioneering HERMES and COMPASS measurements in polarized $ep$ scattering has been given
in the presentations by Riedl \cite{Riedl} and Sozzi \cite{Sozzi}. Examples are inclusive spin structure functions,
various amplitudes and asymmetries from exclusive samples of deeply virtual Compton scattering,
and transverse spin asymmetries. The interpretation of such data is expected to lead to an
improved understanding of the proton spin puzzle, of the related questions of orbital angular momentum
carried by quarks and of the correlations among spin and transverse momentum, as well as 
between longitudinal momentum and transverse position, captured by transverse-momentum
dependent or generalized parton distribution functions.

Related to the basic understanding of the proton structure and of proton scattering cross sections are
studies of the underlying event, multi-parton interactions, particle correlations and diffractive interactions \cite{Pilkington}. 
Recent measurements of this kind are important input to the tuning of Monte Carlo generators.
A fundamental observable is the total inelastic proton-proton cross section, with several
new measurements from ATLAS, CMS and TOTEM. Typically they are presented for 
the fiducial acceptance region and extrapolated to full phase space, as well as differential in
the forward rapidity gap size. In this context, Garcia-Gamez \cite{Garcia-Gamez} has shown 
a very interesting comparison of the LHC results with
an interpretation of shower observables from Auger data, see Fig.\ \ref{fig:proton}, right.

%%%%
%%%%%%%%%%%%%%%%%%%%%%%%%%%% Heavy Flavours
%%%%
\section{Heavy Flavour Physics}
 \label{sec:HF}

Heavy flavour physics as discussed at this conference can be divided into
the following classes: (i) studies of quarkonia systems, (ii) production of heavy flavors
at colliders and (iii) studies of the CKM matrix, CP violation and indirect searches
for new physics with heavy flavor hadrons. 

Addressing the first class, we have seen a large amount of new studies of
charmonium and bottomonium states, with data from BESIII, BELLE, BABAR and the
LHC experiments 
%\cite{Bhardwaj,Guttmann,Wang,Zhao,Tatishvili}. 
\cite{Bhardwaj}$^{-}$\cite{Tatishvili}. 
In particular, most of the
efforts are concentrating on understanding and deciphering the origin of already known or
completely new resonant states, such as 
$X(1835)$, $X(1870)$, $X(2120)$, $X(2370)$, $X(3815)$, $X(3872)$, $X(3823)$ and $\Upsilon(5S)$.
The main questions still to be answered in a satisfactory manner are if these (or some of these)
are indeed tetra-quark states, and/or (loosely-bound) "molecules" of meson-meson pairs,
e.g.\ $D-D^*$ or $B-B^*$. 

A review of heavy flavour production results from the LHC \cite{Aguilo,Kozlinskiy} reveals that, overall,
perturbative QCD gives a rather satisfactory description, with still some discrepancies seen
for particular phase space regions of $p_T$ and/or rapidity distributions. Indeed, such measurements
have been carried out for inclusive open $b$ production, $B$ hadron production as well as
$b$-jet production. Furthermore, angular correlations in events with two $B$-tags have shown
some need for improvements in the Monte Carlo modeling of gluon splitting into $b$ quarks.
Highlights at this conference comprise new results from CMS on $\Lambda_b$  production,
showing a steeper $p_T$ spectrum than observed for $B$ mesons, the first particle discovered
at the LHC, namely the $\chi_b(3P)$ state, an observation by ATLAS now confirmed by D\O, as 
well as new LHCb measurements \cite{Potterat} of $\chi_c$, $\psi(2s)$ and double charm production. Interestingly,
the latter represents a very stringent test for models of double parton scattering.

An excellent review on probing new physics with heavy flavours, and the current experimental
status, was given by Schopper \cite{Schopper}. These efforts can be subdivided in (i) attempts to
constrain the CKM parameters, (ii) measurements of direct or mixing-induced CP violation and (iii) the
searches for very rare decays. Some of the most important new results or updates presented
at the conference, concerning these areas, are LHCb studies of direct CP violation in hadronic 
B decays \cite{Tilburg,CJones} ($B\rightarrow hh'$), new results on CP violation in $B_s$ mixing \cite{Dorigo},
a large number of new results on rare decays \cite{Dorigo,Smizanska,Parkinson}, such as $B_s\rightarrow\mu^+\mu^-$ from 
LHCb, CMS, ATLAS and CDF, as well as $B\rightarrow K^*\mu^+\mu^-$ and further rare decays from
LHCb, e.g.\ $B^+\rightarrow \pi^+\mu^+\mu^-$ and $B\rightarrow 4\mu$. In the case of  $B\rightarrow K^*\mu^+\mu^-$,
LHCb has presented the world's first measurement of the zero crossing point for the forward-backward asymmetry, giving nice
agreement with the SM prediction. Finally, new results on the CP-asymmetry in charm have been presented \cite{Dorigo},
confirming values of this asymmetry around the $-1\%$ level and at $\sim4 \sigma$ from zero, thus indicating a larger
value than expected from some of the currently available SM estimates.

\begin{figure}[htb]
\centering
\begin{tabular}{lr}
\includegraphics[width=6cm]{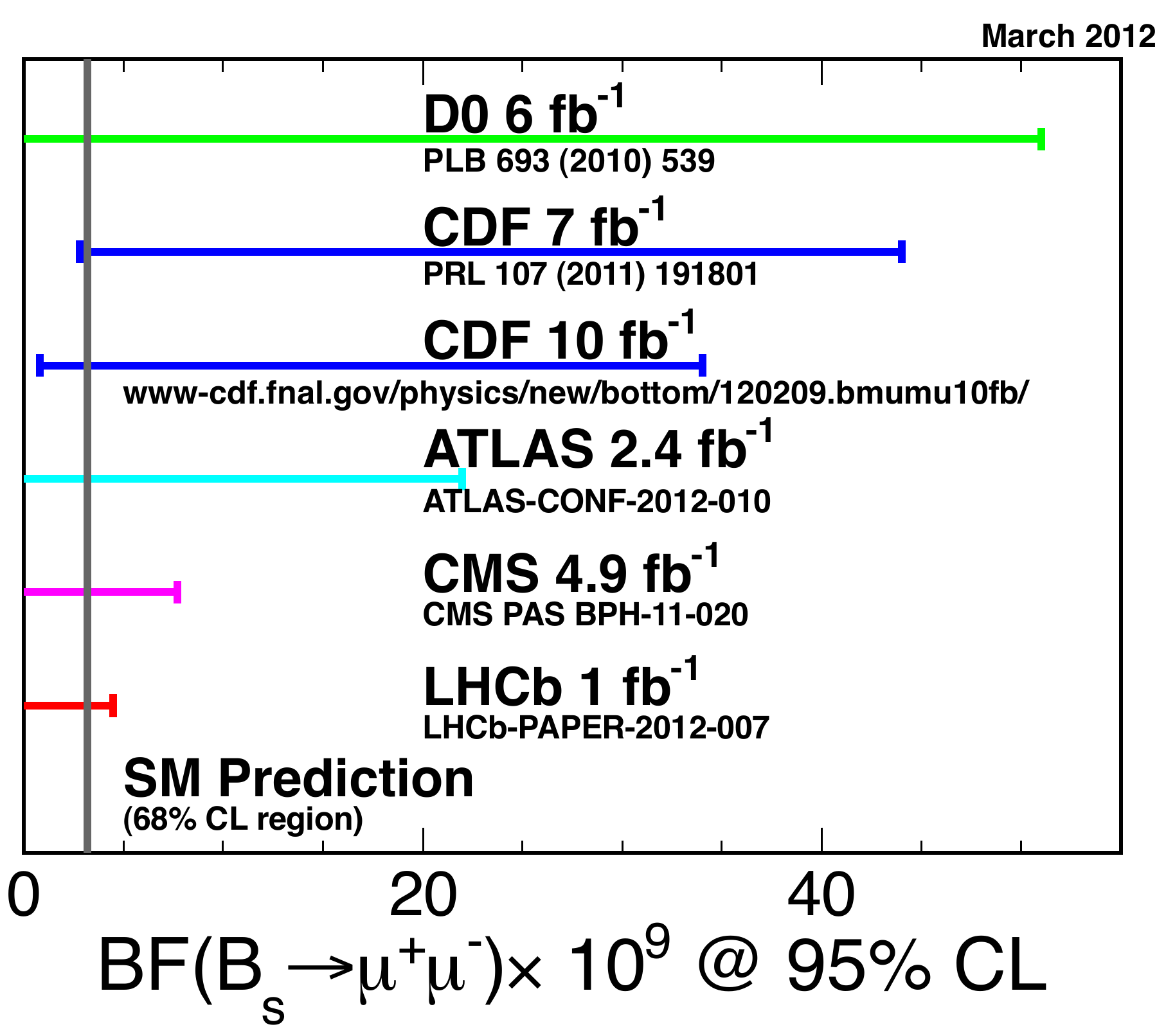} &
\includegraphics[width=7.5cm]{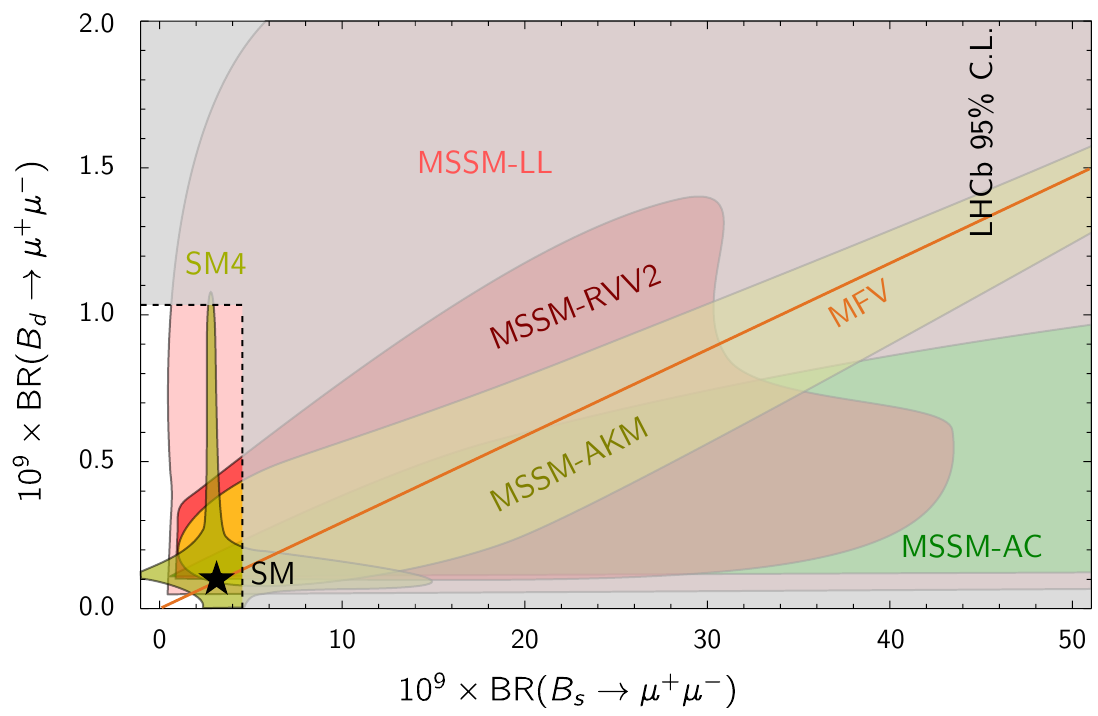} 
\end{tabular}
\caption{Left: Summary of upper limits on the branching ratio for $B_s\rightarrow\mu^+\mu^-$;
             Right: Impact of these limits on the parameter space of SUSY models (from D. Straub, Moriond EWK 2012).}
\label{fig:Bsmumu}
\end{figure}

The recent progress towards identification of rare decays attracts most of the current attention.
Figure \ref{fig:Bsmumu} (left) gives a summary of the most recent upper limits obtained on the 
branching ratio for $B_s\rightarrow\mu^+\mu^-$, which is 
very sensitive to contributions from new physics such as SUSY
and predicted by the SM to be $(3.2\pm0.2)\times10^{-9}$.
The current world's best limit, obtained by LHCb from their 1 fb$^{-1}$ dataset, is
$BR(B_s\rightarrow\mu^+\mu^-) < 4.5\times10^{-9}$, closely followed by CMS which finds
$BR(B_s\rightarrow\mu^+\mu^-) < 7.7\times10^{-9}$ from their full 2011 dataset (5 fb$^{-1}$). Also ATLAS has presented
a first limit from an analysis of a 2.4 fb$^{-1}$ data sample, 
and CDF has shown an update based on their full RunII statistics. Their new result
does not further enhance, but rather reduce, a slight excess found in their 7 fb$^{-1}$ sample.
The strong power of this observable is nicely illustrated by Fig.\ \ref{fig:Bsmumu} (right) and further
discussed in Sec.\ \ref{sec:NewPhen} below, showing
that these recent results exclude a very large portion of parameter space for SUSY models \cite{Straub}.

In conclusion, the results on heavy flavour physics presented at this conference could be summarized by
naming LHCb as an "anomaly terminator". This is because (i) earlier indications of a large phase $\Phi_s$ 
in $B_s$ mixing have not been confirmed, the current results showing nice agreement with SM expectations; 
(ii) the measured forward-backward asymmetry and derived parameters in the $B\rightarrow K^*\mu^+\mu^-$ decay
also agree with the SM, and thus do not confirm earlier hopes of possible signs of new physics in this decay;
(iii) and finally the limit on the $B_s\rightarrow\mu^+\mu^-$ branching ratio is approaching the SM value,
with a first measurement to be expected later in 2012. Nevertheless, for those believing in new physics showing
up in heavy flavour systems, their is now some new hope due to the large CP-asymmetry found in charm.
However, care should be taken here, since SM predictions in this area suffer from large long-distance (non-perturbative)
QCD effects, thus are notoriously difficult to predict, i.e.\ in the end it could simply turn out that the observed
large asymmetry could be ascribed to such QCD effects. Overall, the phenomenologists are more and more given
a fantastic set of data and experimental constraints, which allow putting strong limits on new physics, in particular
when combined with other observables, such as direct searches at colliders (see below).

%%%%
%%%%%%%%%%%%%%%%%%%%%%%%%%%% Hard QCD
%%%%
\section{Tests of perturbative QCD}
 \label{sec:QCD}

Measurements of hard-scattering cross sections, with jets, photons or
vector bosons in the final state, are interesting because of several reasons:
(i) it allows probing higher-order predictions of perturbative QCD for the
hard-scattering part of the overall process; (ii) parton distribution functions
can be constrained; (iii) SM predictions can be tested, in particular QCD calculations,
as implemented in various codes and MC generators, for processes
which are important backgrounds for new physics searches. At this conference
a large number of new results in these directions have been presented, in general
showing a remarkable agreement of theory and data. We note in passing that
a more extensive review of this subject has been published recently \cite{Butterworth:2012fj}.

A central component of those measurements, which contain jets in the final state,
is the excellent control of the systematic uncertainty due to the jet energy scale. This is
essential because of the nature of the steeply falling cross sections as a function of
the jet $p_T$. By now the LHC experiments master this effect already at a remarkable 
level of precision, e.g., around 2\% or even better for central jets and a $p_T$ range of
about 50 to several hundred GeV.

\begin{figure}[htb]
\centering
\begin{tabular}{lr}
\includegraphics[width=7.5cm]{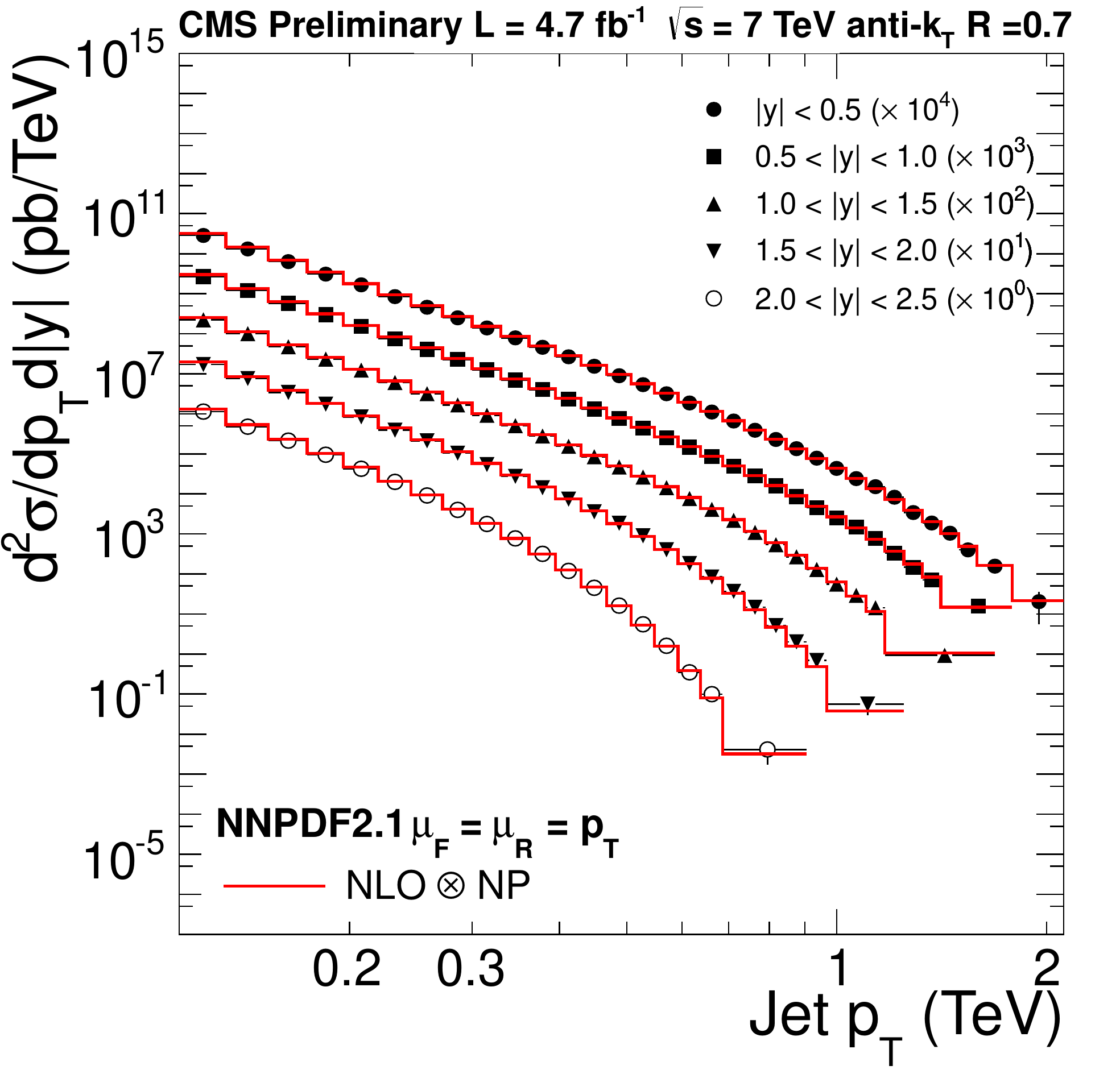} &
\includegraphics[width=7.5cm]{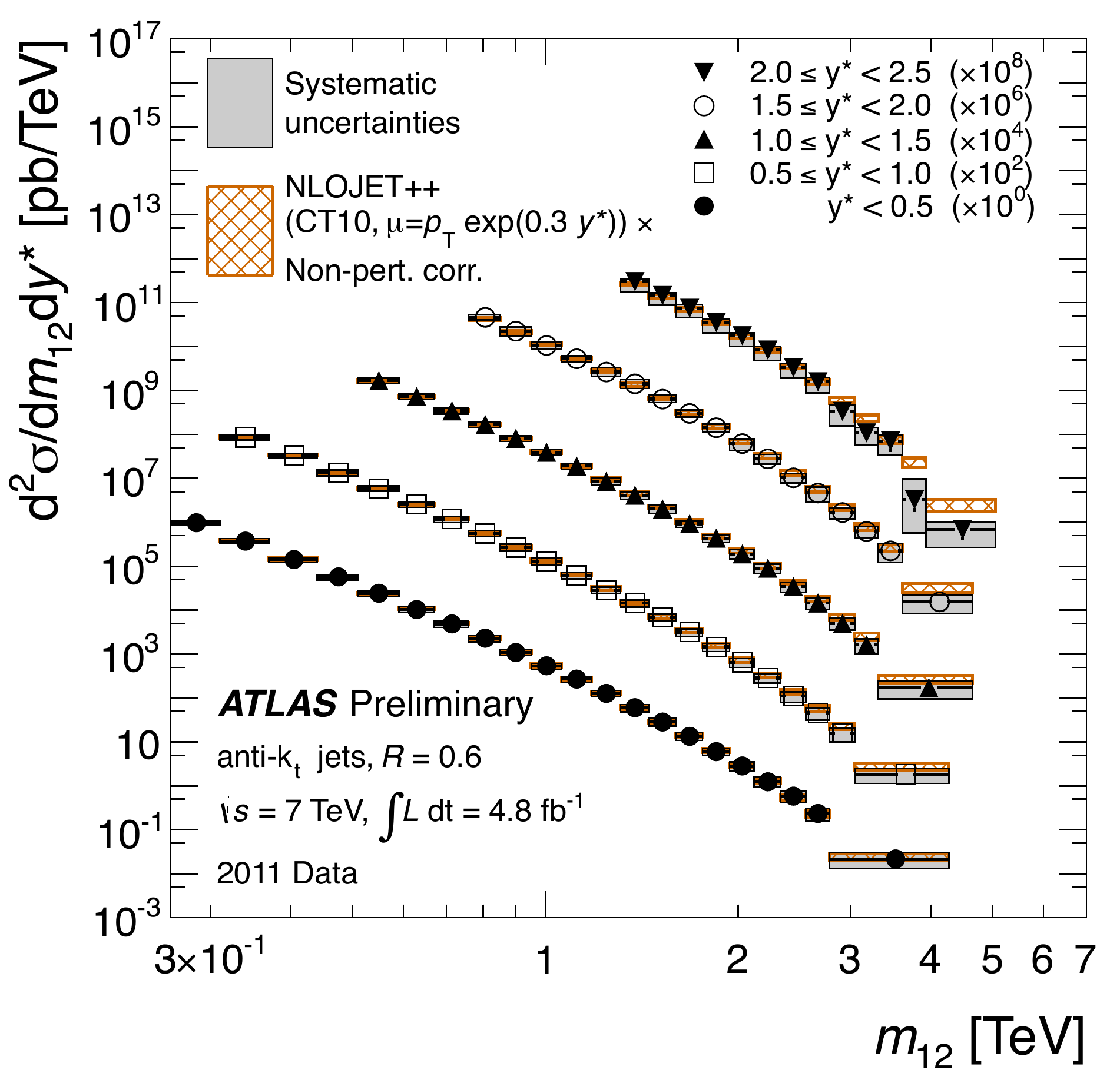} 
\end{tabular}
\caption{Left: Inclusive jet production, as a function of jet $p_T$ and rapidity, measured by CMS; 
%              CMS-PAS-QCD-11-004;
	 Right: ATLAS data on dijet production.
%             ATLAS-CONF-2012-021.
	}
\label{fig:incljets}
\end{figure}

Concerning jet production at the LHC, new results have been presented \cite{Jones} for
inclusive jet production, dijet production as a function of dijet invariant mass and
jet rapidity separation, as well as third-jet activities. In particular, new measurements have appeared
on the inclusive jet cross section as a function of jet $p_T$ by CMS, 
and dijet production by ATLAS, based on the full
2011 dataset, cf.\ Fig.\ \ref{fig:incljets}. Overall, the agreement of next-to-leading order (NLO)
QCD predictions with data over many orders of magnitude is rather impressive. The inclusive
jet cross section has been compared to predictions based on a large set of pdfs, showing in general
good agreement within theoretical and experimental uncertainties. In the dijet case, where
the data have an impressive reach up to about 4 TeV in dijet mass, some discrepancies are found
at very large masses and large dijet rapidity separation, a region where NLO predictions probably
reach their limit of applicability. A similar observation is made by a dedicated CMS analysis, which
studied central jet production with the additional requirement of a second jet in the forward region.
They found some significant disagreements among data and MC models. Finally,
ATLAS has presented a measurement of the $D^*$ fragmentation function, showing a sizable
discrepancy, with MC clearly underestimating the yield in the data. This might point to a problem
with the simulations for gluon splitting to charm, similarly to the observations for the b-quark case in an earlier CMS
measurement of $B\bar{B}$ angular correlations \cite{Khachatryan:2011wq}.

\begin{figure}[htb]
\centering
\begin{tabular}{lr}
\includegraphics[width=7.5cm]{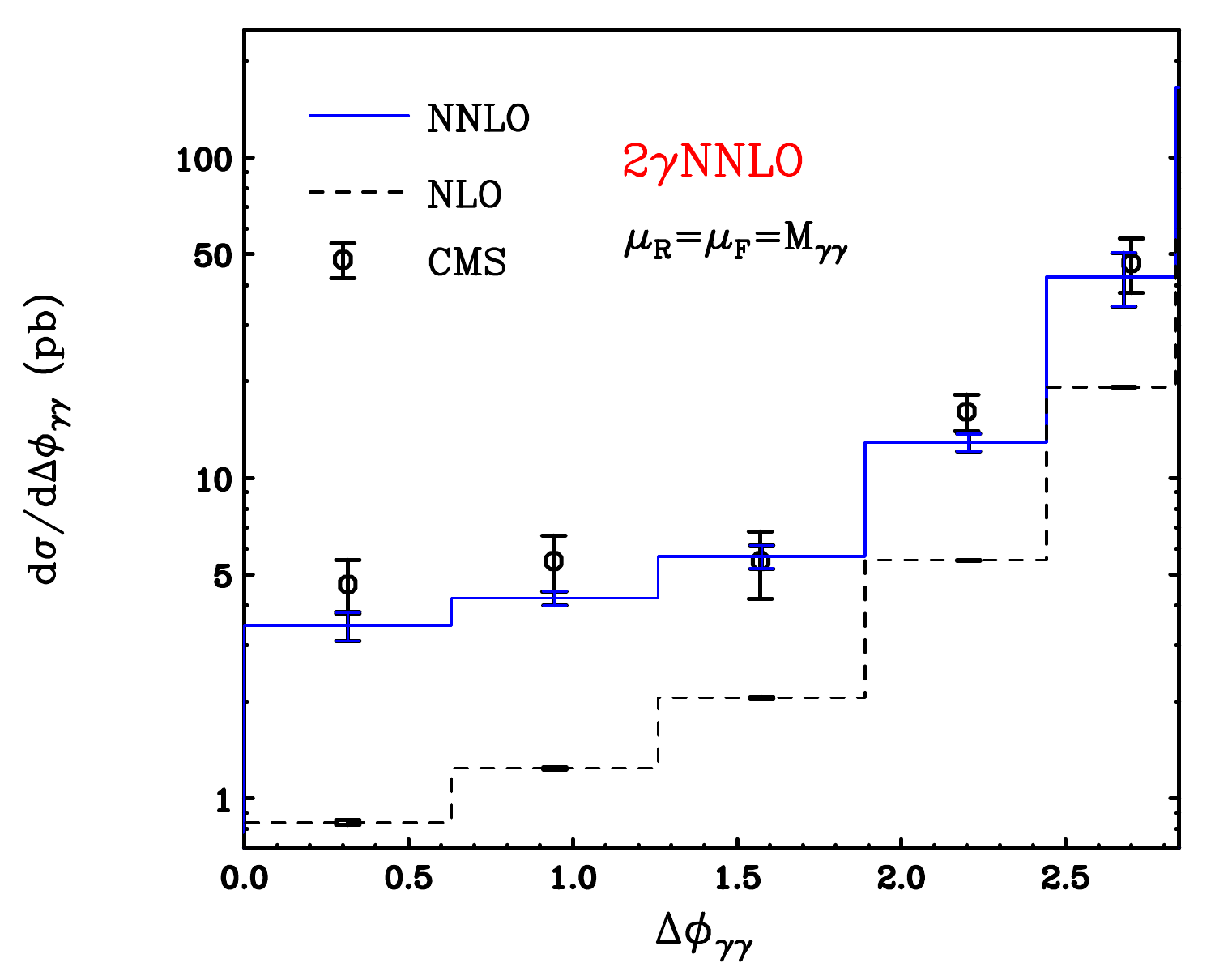} &
\includegraphics[width=6.8cm]{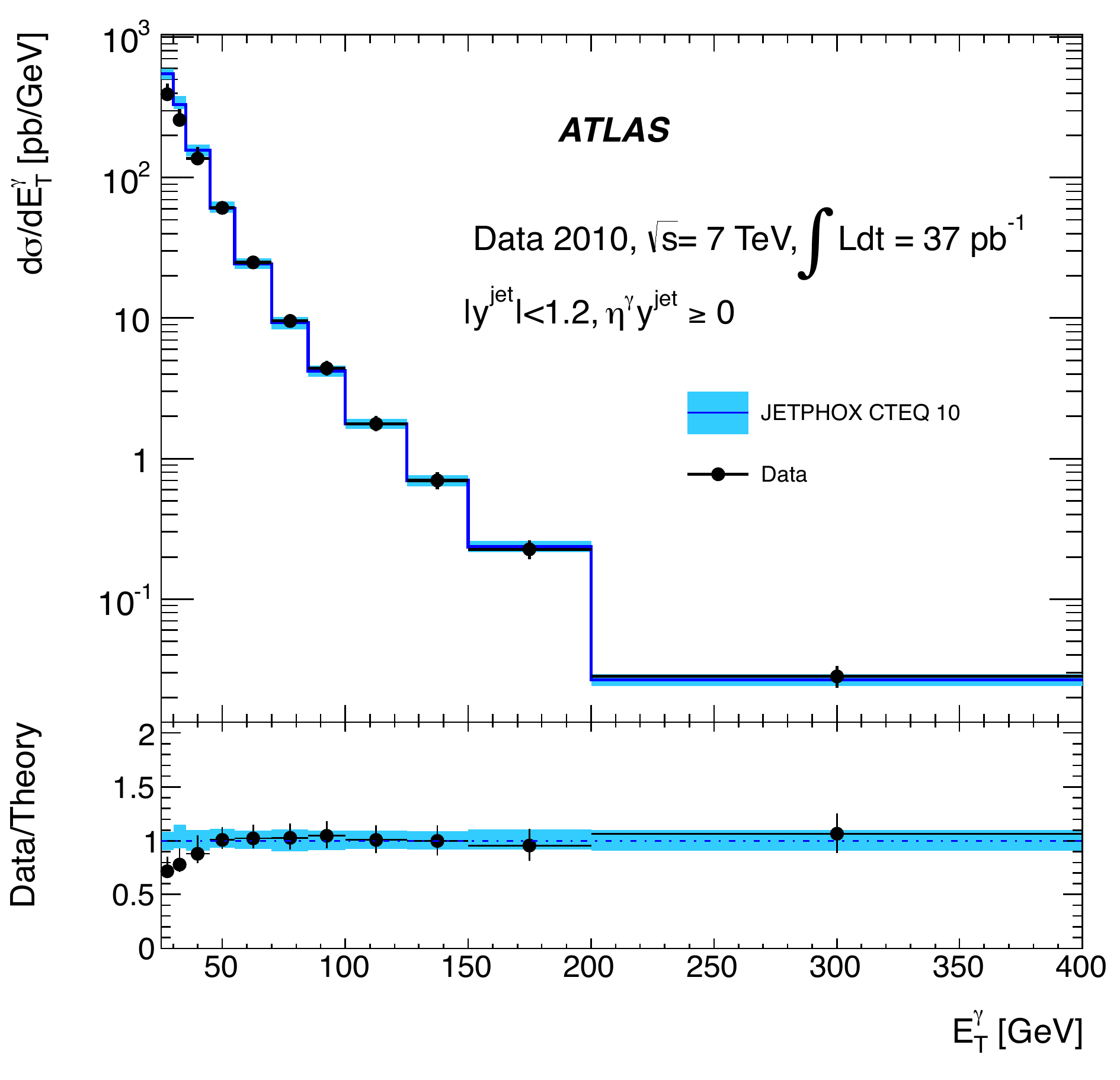} 
\end{tabular}
\caption{Left: Comparison of CMS data and QCD predictions for the di-photon azimuthal separation;
	      Right: photon plus jet production cross section from ATLAS. 
%             ATLAS arXiv:1203.3161
	}
\label{fig:photons}
\end{figure}

New results on inclusive photon, di-photon and photon plus jet production
at the TEVATRON and the LHC have been presented by
Dittmann \cite{Dittmann} and Gascon-Shotkin \cite{Gascon-Shotkin}.
Among the highlights of this year, there is a new calculation \cite{Catani:2011qz} at next-to-NLO (NNLO) level for
di-photon production, which finally brings the theory into agreement with data in the region
of small azimuthal separation (Fig.\ \ref{fig:photons}, left). In that region of phase space the previously
available NLO calculation is effectively a leading order approximation, which underestimates
the data obtained for this distribution both at the LHC and the TEVATRON. Thus here we have
a spectacular example for the need of NNLO calculations, for the description of particular
variables in specific regions of phase space, not only because of radiative corrections, but also
because of the appearance of new partonic channels in the initial state only at a certain order
of perturbation theory. Also worth mentioning is the first LHC measurement on photon plus jet
production by ATLAS, as a function of several kinematic variables and differential in the
photon-jet angular separation. This is a classical study for hadron colliders, in particular because
of the sensitivity to the gluon pdf. The data are in good agreement with NLO predictions (Fig.\ \ref{fig:photons}, right), besides
some deviations seen for photon $p_T$ below 50 GeV. A similar observation had been made
for inclusive photon production. Also worth mentioning is a measurement of angular decorrelations
in photon plus 2 or 3 jets final states by D\O, showing nice evidence for the need to include double parton
scattering contributions into the theoretical predictions.

\begin{figure}[htb]
\centering
\begin{tabular}{lr}
\includegraphics[width=6cm]{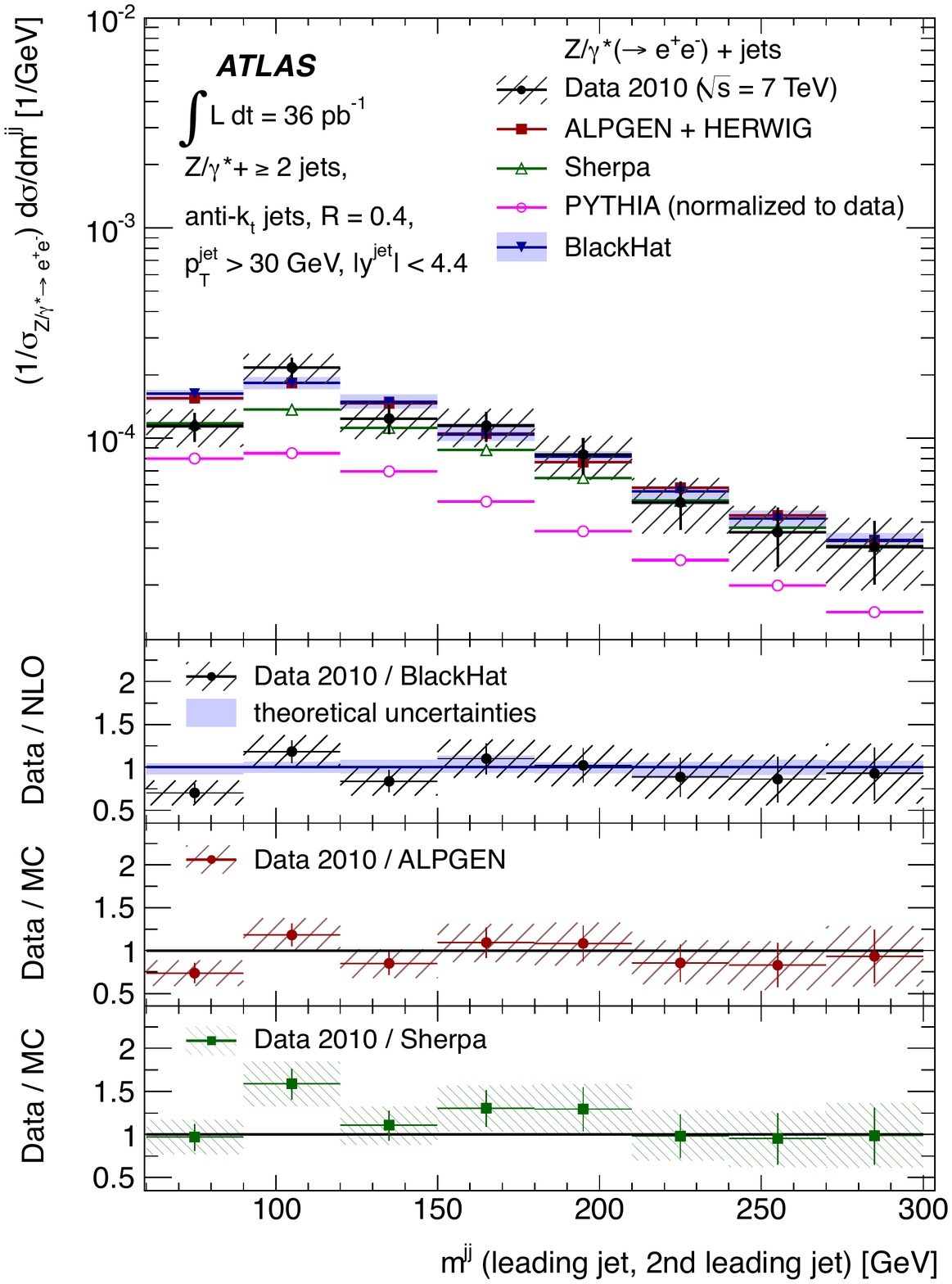} &
\includegraphics[width=9cm]{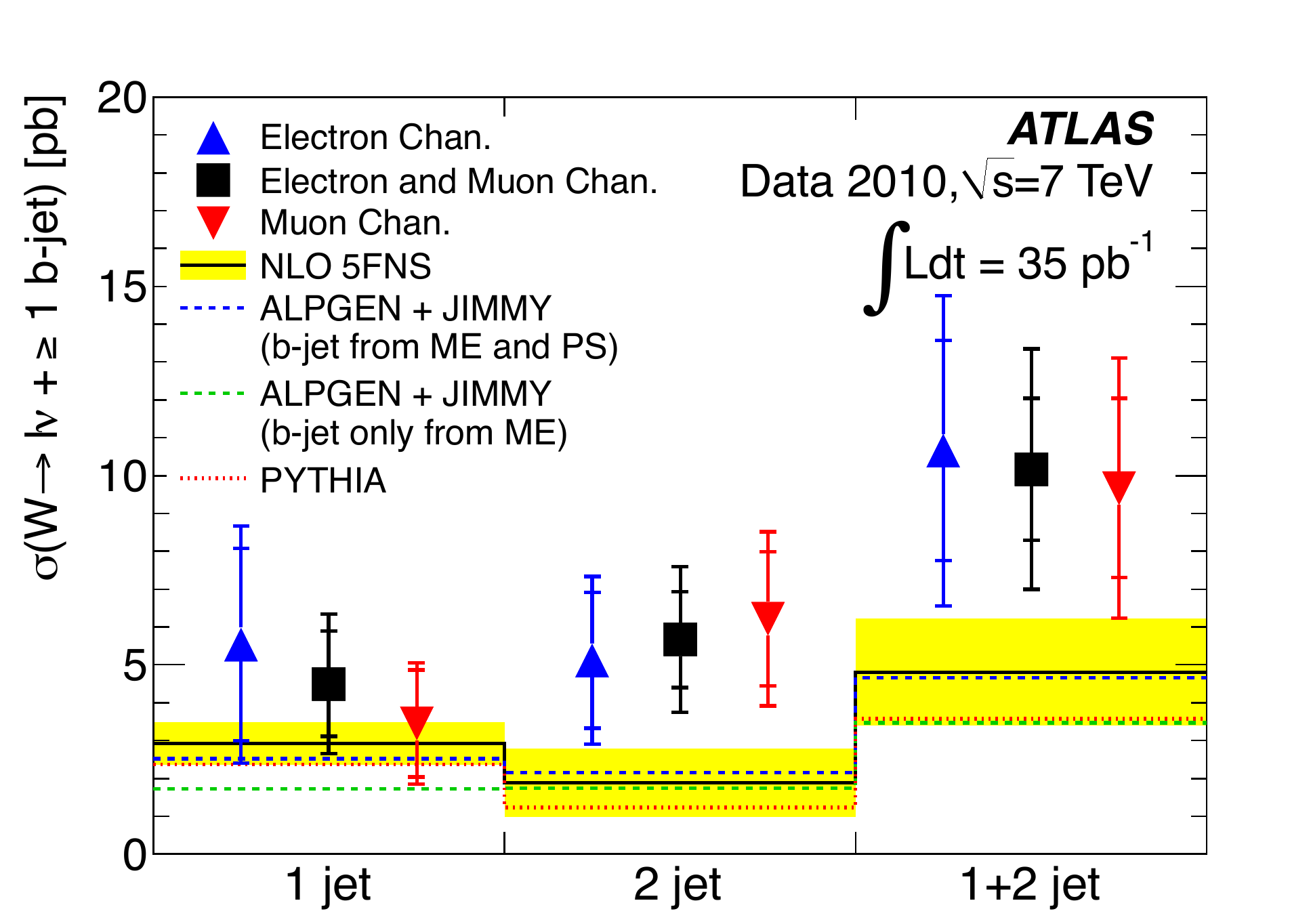} 
\end{tabular}
\caption{Left: Invariant mass of the leading and 2nd leading jet, measured by ATLAS in $Z+$jets final states;
%ATLAS arXiv:1111.2690
              Right: ATLAS results on the production of a $W$ boson in association with $b$-jets.
%             ATLAS PLB 707 (2012) 418
	}
\label{fig:Vjets}
\end{figure}

Whereas the excellent agreement of data with NNLO QCD predictions, for the inclusive production
of $W$ and $Z$ bosons, had already been shown and discussed at earlier conferences, this year special
focus has been put on the study of vector boson plus jet production \cite{Bandurin,Paramonov}. These
processes are extremely important backgrounds for searches of supersymmetry and the Higgs,
especially for associated Higgs production in the low mass region. Furthermore, such measurements
allow for testing different approaches to the implementation of perturbative QCD calculations into MC codes,
such as at fixed order (NLO) or based on the matching of leading order matrix elements with parton showers,
for example in MADGRAPH, ALPGEN or SHERPA. Thanks to important recent advances,
NLO calculations are now available up to high jet multiplicities \cite{Soper}.
Concerning such jet multiplicities in $W$ (or $Z$) plus jet production, as well as angular correlations among the jets,
overall a very good agreement with the NLO and matched calculations is found. Also dijet masses and
the $H_T$ distribution (scalar sum of jet momenta) are well modeled over large regions of
phase space, where the various calculations are applicable (Fig.\ \ref{fig:Vjets}, left). For the more
specific case of vector boson plus heavy flavor production ($b$- or $c$-tagged jets), a rather consistent picture
seems to appear from the TEVATRON and the LHC: data and NLO QCD predictions agree, within the
sometimes sizable theoretical and experimental uncertainties, for $W+c$ and $Z+b$ production, whereas
deviations are found for $W+b(b)$  (Fig.\ \ref{fig:Vjets}, right). This is interesting, again also because of the
relevance of this process for the Higgs search. Finally, first studies of angular correlations in $Z+bb$ final
states have been presented by CMS.

Going lower in production cross section for electro-weak particles, the most relevant and often studied
processes are di-boson production ($W\gamma, Z\gamma$, WW, WZ, ZZ), for various decay channels
of the vector bosons. An interesting new measurement of $VZ(\rightarrow b\bar b)$ production at the TEVATRON
\cite{Vizan} is further discussed in section \ref{sec:Higgs} below. The large and by now rather complete
set of LHC results is summarized in more detail in Ref.\ \cite{Malcles}. The picture arising is that all the
aforementioned processes, measured with statistics up to 5 fb$^{-1}$, are in agreement with NLO QCD
predictions, which then allows to put stringent constraints on anomalous trilinear gauge couplings. It has
been remarked that the measured $WW$ cross section appears to be slightly higher (however, not
at a statistically significant level) both in ATLAS and CMS,
compared to the NLO predictions. Since this process is particularly relevant for the understanding of
electro-weak symmetry breaking, it will be interesting to follow up on future results in this area.
In the past, a bump in the dijet mass distribution for $W+2$ jets production, observed by CDF, had caused
a certain amount of excitement. However, at this conference both D\O\ and CMS presented results, which do not
confirm that finding. Finally, an interesting new LHC measurement, related to the $ZZ$ and $H\rightarrow ZZ$ processes,
has been put forward by CMS, namely  the first observation at a hadron collider of $Z\rightarrow 4 \ell$. While interesting
in itself, this process will turn out to be an extremely useful standard candle for controlling the absolute
mass scale, the mass resolution and the reconstruction efficiencies for the Higgs search in the four-leptons channel.

We close this section on tests of perturbative QCD by mentioning a nice re-analysis of JADE data for the 3-jet rate, 
used to precisely determine the strong coupling constant at NNLO+NLLA approximation \cite{Kluth}. Indeed, it 
is shown that this measurement has an uncertainty due to higher order QCD corrections below the 1\% level, 
and is dominated by hadronization model systematics. Similarly, new recent results were also shown on
jet production and $\alpha_s$ determinations based on HERA data \cite{Kapishin}.

%%%%
%%%%%%%%%%%%%%%%%%%%%%%%%%%% TOP
%%%%
\section{Physics of the Top Quark}
 \label{sec:TOP}

The top quark is given special attention because of several reasons: it is by far the heaviest of all
quarks, and with a mass of the order of the electro-weak scale it is conceivable that the top plays
a special role in electro-weak symmetry breaking. Furthermore, it is considered to be a possibly
important gateway to new physics. Until recently the TEVATRON has been the only player in the field.
However, the LHC has quickly risen to the status of a ''top factory" and
the LHC experiments start to play the leading role more and more. A central test of 
SM predictions is the measurement of the top-pair production cross section. The LHC experiments
have presented new results \cite{Aracena} for a large number of channels (leptons+jets, dileptons, $\tau+\mu$,  
$\tau+$jets, all hadronic), analyzing data sets between 0.7 and 4.7 fb$^{-1}$. The currently combined best
cross section values found by ATLAS and CMS are 
$\sigma_{t\bar t} = 177 \pm 3\mathrm{(stat})^{+8}_{-7}\mathrm{(syst)} \pm 7\mathrm{(lumi)}\,\mathrm{pb}$ and
$\sigma_{t\bar t} = 165.8 \pm 2.2\mathrm{(stat)}\pm 10.6\mathrm{(syst)} \pm 7.8\mathrm{(lumi)}\,\mathrm{pb}$, respectively.
Here one should highlight that the experimental uncertainty has already achieved a level of 6\%, which is
smaller than the uncertainty on the theoretical predictions. It would be interesting to see an ATLAS-CMS combination,
also in light of the very slight tension which appears from these two experimental results and the fact that 
most likely part of the systematic uncertainties are correlated. Nevertheless, both results are in agreement
with expectations from perturbative QCD, and one should start considering the possible impact of this
cross section on pdf determinations \cite{Schwinn}.

\begin{figure}[htb]
\centering
\begin{tabular}{lr}
\includegraphics[width=7.5cm]{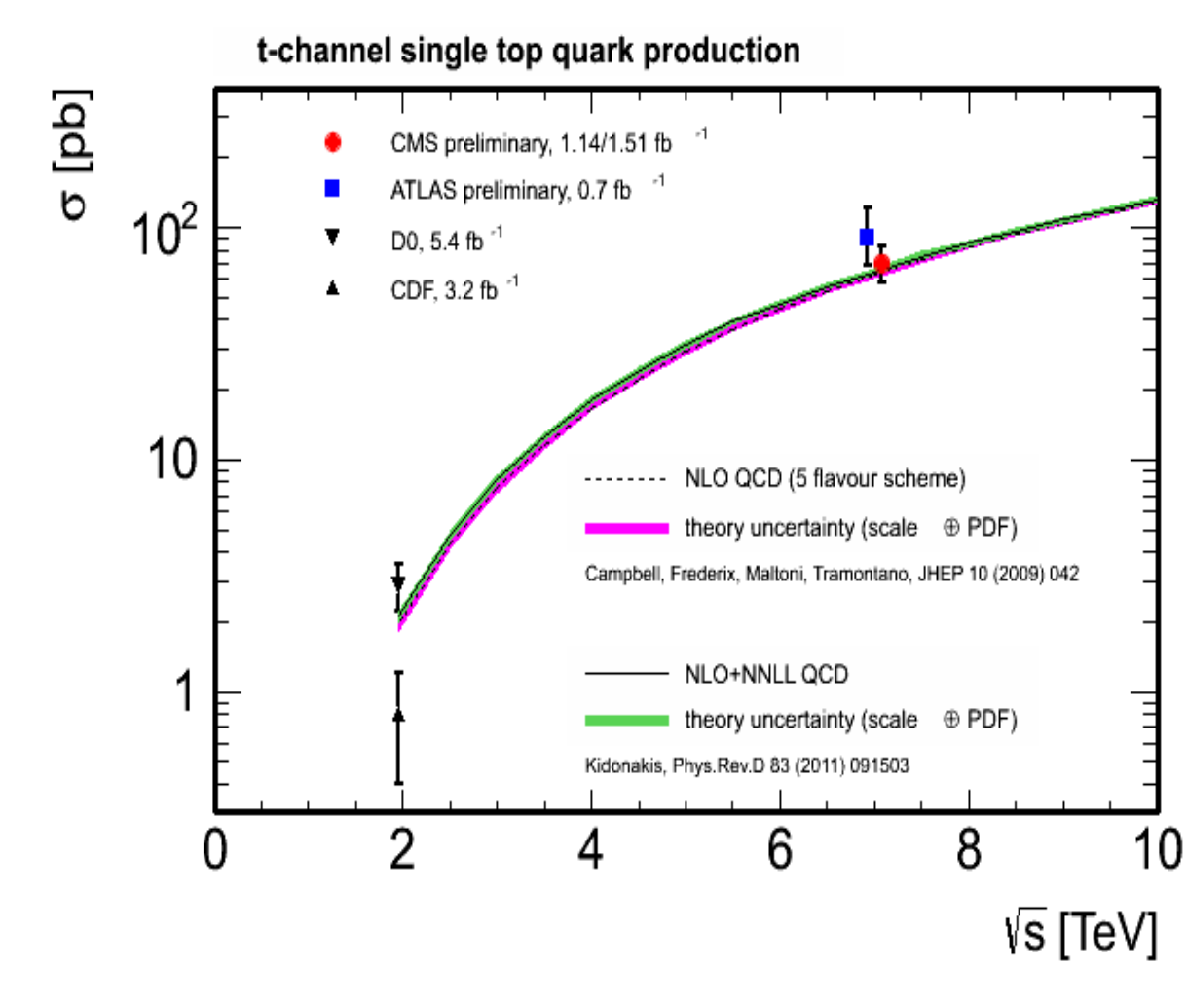} &
\includegraphics[width=6.5cm]{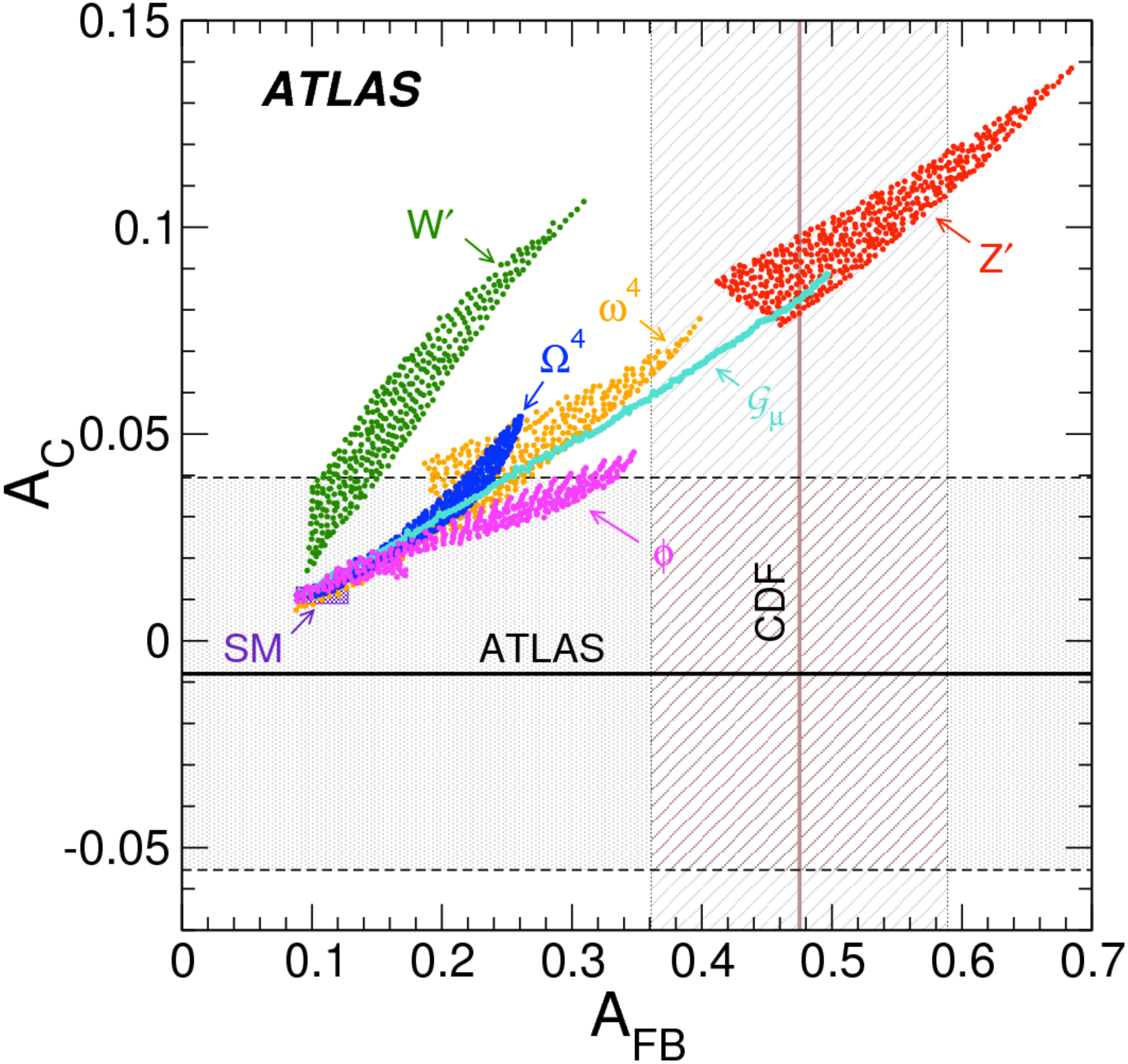} 
\end{tabular}
\caption{Left: Single top production cross section as a function of centre-of-mass energy;
              Right: Comparison of top asymmetry measurements by ATLAS and CDF, and interpretation
                in terms of new physics models.
%             ATLAS arXiv:1203.4211
	}
\label{fig:top}
\end{figure}

The studies of single top production are steadily progressing, both at the TEVATRON \cite{Wu} and the LHC \cite{Suarez}.
Thanks to the considerably enhanced cross section at the LHC compared to the TEVATRON,
ATLAS and CMS have already reached an accuracy of $\sim20\%$ in the measurement of $t$-channel production (Fig.\ \ref{fig:top}, left).
A clear wish has been expressed at the conference for harmonizing, among the LHC experiments, the treatment
of theoretical uncertainties in this class of measurements. CMS has interpreted the cross section measurement
in terms of $|V_{tb}|$ and extracted a measurement at the 10\% accuracy level. Besides this production channel,
a considerable effort is spent by all experiments, in order to close in on the $tW$ and $s$ channels.

What concerns the top mass, the TEVATRON is still leading, with the world's most precise measurement, from
a TEVATRON combination, presented \cite{Brandt} to
be $m_t = 173.2 \pm 0.6\mathrm{(stat)}\pm 0.8\mathrm{(syst)}$ GeV, noteworthy a quark mass measurement with
a relative uncertainty of 0.54\%. Further improvements are still expected until the final analysis of the full Run II dataset. 
However, the LHC is catching up. For example, CMS has come up \cite{Blyweert} with their latest best result of
$m_t = 172.6 \pm 0.6\mathrm{(stat)}\pm 1.2\mathrm{(syst)}$ GeV, thus already achieving the same statistical precision
as the TEVATRON experiments. However, it was noted that this determination does not yet consider some
systematic uncertainties, such as color reconnection and underlying event effects. 
There is certainly an interest in 
obtaining an LHC, and ultimately an LHC-TEVATRON, combination for this important parameter. Such an effort should
then also help in synchronizing the treatment of systematic effects by the different experiments.
A further observation made
at the conference was that all experiments use the $W$ mass as a kinematic constraint in their analyses, meaning
that there is some correlation between the top and $W$ mass measurements. On the other hand,
in the electro-weak precision tests, where the consistency among $m_t, m_W$ and $m_H$ is tested (see also below), 
such a correlation is not taken into account. However, because of the largely different levels of precision achieved for
these mass determinations, in the end this is not a serious issue, most likely.  A somewhat "disturbing" aspect
of the direct top mass determinations from kinematic reconstruction is the not really well defined
meaning of the finally extracted parameter. While it is supposed to be close to a definition according to a pole-mass scheme,
currently a theoretically sound understanding is not available, which triggers some experts to question if we really know this
quark mass at the 0.5\% accuracy level \cite{Soper}. On the other hand, a theoretically very well defined approach is given by the
extraction of the top mass (typically in the form of a running mass) from a top cross section measurement. In view of the
ever improving precision on the latter (see above), this becomes more and more interesting. So far an accuracy of
$\mathcal{O}(7$ GeV) is attained, mostly dominated by pdf uncertainties, and achieving a 5 GeV error seems to be viable \cite{Schwinn}.

Besides production cross sections and mass, an amazing amount of further top properties have been studied
by the TEVATRON and LHC experiments \cite{Lister,Mietlicki}. These comprise spin correlations, $W$ helicity and polarization
in top decays, extractions of $|V_{tb}|$, the top width, $m_t - m_{\bar t}$, the electric charge of the top, the charge asymmetry,
searches for anomalous couplings and flavour-changing neutral currents, as well as  a first study of jet veto effects in top-pair production.
Basically for all these properties and observables agreement is found among data and SM predictions, besides the
well-known discrepancy found for the forward-backward asymmetry ($A_{\mathrm{FB}}$) at the TEVATRON.
CDF has presented a differential study of $A_{\mathrm{FB}}$ as a function of the invariant mass of the $t\bar t$ system, showing
a steeper slope in data compared to theory. Probably further improved understanding, eg., of
non-perturbative effects when correcting from particle to parton level, as well as from higher-order QCD, is needed before
establishing this as a significant hint for new physics. This cautious approach appears to be supported by an interesting
interpretation by ATLAS of their latest measurement of the top charge asymmetry, $A_{\mathrm{C}}$. When comparing
their measurement, as well as CDF's $A_{\mathrm{FB}}$ result, to predictions for $m_{t\bar t} > 450$ GeV, there appears to arise 
some tension, Fig.\ \ref{fig:top}, right. For example, while some physics beyond the SM, such as a heavy Z boson, would still be
consistent with the CDF measurement, its effect would lead to a larger $A_{\mathrm{C}}$ than observed by ATLAS.

%%%%
%%%%%%%%%%%%%%%%%%%%%%%%%%%% New Phenomena
%%%%
\section{Searches for New Phenomena}
 \label{sec:NewPhen}

 The searches for new physics, now dominated by the LHC results, can be roughly classified
 into two large sectors, namely (i) those concentrating on signatures of SUSY particles, and (ii)
the large class of searches for other particles and interactions beyond the SM. 
The sheer amount of SUSY exclusion plots shown at this conference is testimony of the enormous
efforts invested at the collider experiments, in order to get any hint of SUSY components in the data.
Typical classifications of the analyses follow topological considerations, such as looking for events with large
missing transverse energy (MET), due to the possible production of weakly interacting massive SUSY particles,
accompanied by high-$p_T$ jets, one or two opposite or same-sign leptons, more than two leptons or photons.
The interpretation of the, so far unsuccessful, searches of any deviation from the SM predictions, is carried
out in various manners; either in the context of since long established specific SUSY incarnations, with very constrained
parameter sets, such as mSUGRA or cMSSM, or in a more general approach as implemented in so-called
Simplified Models (see e.g.\ Ref.\cite{Alves:2011wf}). 
In this case basic properties of particle cascades, arising from the decays of heavy particles
such as pair-produced gluinos, are explored. At the conference first results were presented based on the full
2011 statistics, showing the potential for big advances in terms of excluded parameter space.
In simple terms, the current results of ''generic'' squark and gluino searches, in the topologies as mentioned before, 
allow setting limits around the TeV scale, if interpreted in scenarios such as the cMSSM \cite{Duflot}. Thus, with 
the first two years of LHC data this mass scale is pushed rather high, such that some start to consider giving up (at least to
some extent) naturalness arguments. On the other hand, first attempts have already started, and will be pursued with
much more vigor in 2012, regarding the searches for third generation squarks. So far limits in those cases are not too strong,
roughly around 300 GeV. Such efforts are, e.g., motivated by models where the first generation squarks are pushed to very
high mass scales, whereas only the third generation is kept light, around the electroweak scale, arguing that after all
naturalness can be maintained if the effects from top loops, which dominate radiative corrections to the Higgs mass, are controlled
by contributions from particles such as stops. These searches could turn out to be rather difficult, in particular 
if the mass separation between the top and third-generation spartners is not too large.
Related to these SUSY searches, there are two further aspects worth mentioning: (i) when looking at the enormous amount
of analyses, in the end always condensed into a few exclusion plots, one easily forgets to appreciate the large ingenuity
and the many new ideas, which are at the basis of those results. In particular, during these last years a large set of
new observables, which are differently sensitive to SM backgrounds and to the appearance of new heavy particles, 
have been established, as well as many clever, so-called data-driven, methods have been developed,
 in order to estimate SM background contributions to the search regions. In this context, also observables are studied,
 such as the ratio of $Z+$jet over $\gamma+$jet production as a function of $H_T$ and/or jet multiplicity, which are interesting
 in itself from a SM point of view.
 
  \begin{figure}[htb]
\centering
\begin{tabular}{lr}
\includegraphics[width=7.5cm]{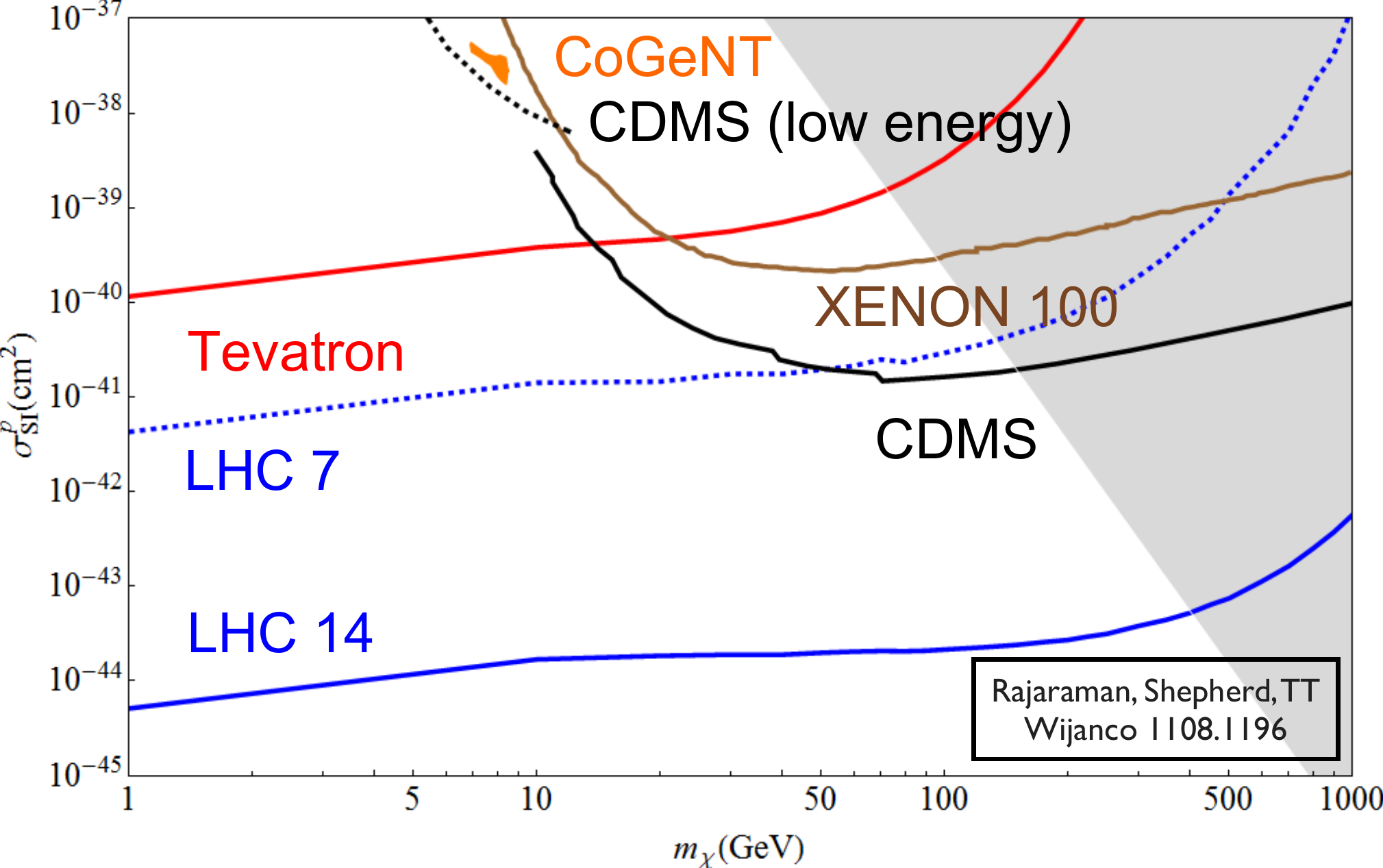} &
\includegraphics[width=6cm]{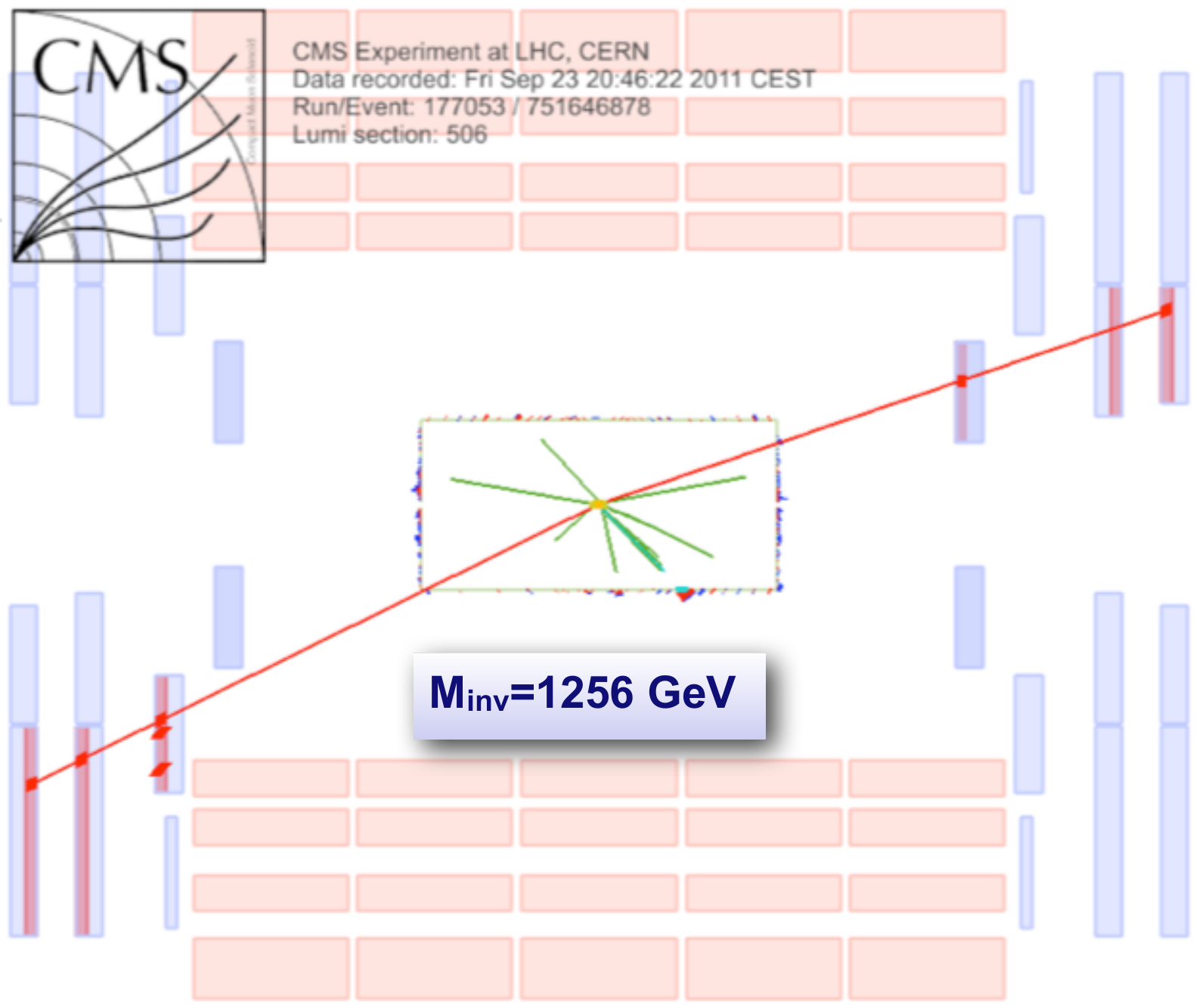} 
\end{tabular}
\caption{Left: Complementarity of the searches for DM candidates at colliders and by direct DM detectors: shown is an 
               example of limits on the spin-independent WIMP-nucleon scattering cross section in an iso-spin violating scenario, 
               with regions above the lines excluded;
               Right: Dimuon event with very large invariant mass as detected by the CMS experiment.}
\label{fig:BSM}
\end{figure}

 The discussions of SUSY searches have focused on two further highly-interesting aspects. Tait \cite{Tait} highlighted
 the important complementarity of searches for dark matter (DM) candidates (in particular Weakly Interacting Massive 
 Particles, WIMPs), as carried out at colliders, with direct DM searches. Whereas at colliders we probe the parton-DM
 couplings, in direct DM searches one explores the coherent nucleon-DM scattering. An advantage of collider searches
 is their reach towards very small DM masses, by e.g.\ looking for monojet signatures induced by direct DM pair-production
 and a jet from initial state radiation \cite{Sonnenschein}. 
 This complementarity is nicely expressed in
 exclusion plots as shown in Fig.\ \ref{fig:BSM}, left. 
  Another example of complementarity was underlined by Mahmoudi \cite{Mahmoudi},
 who analyzed the constraining power, in terms of SUSY models, arising from heavy flavor physics, such as 
 rare decays ($B\rightarrow K^*\mu^+\mu^-, B_s\rightarrow\mu^+\mu^-$) mentioned above, or from searches for 
 (supersymmetric) Higgs bosons. In simple terms, the direct searches push the masses of (first generation) particles 
 higher and higher, and rare decays such as  $B_s\rightarrow\mu^+\mu^-$ strongly constrain $\tan\beta$ to lower values,
 therefore creating tension with other observables such as the muon $g-2$ result. Though, concerning the latter,  
participants at the conference highlighted the need for a still better understanding of the theory uncertainties, before taking
this tension too seriously. Finally, if the current exclusion limits for a very light Higgs below about 120 GeV are taken
at face value, particular implementations of SUSY breaking, such as gauge mediation (GMSB), can be considered to be ruled out.

Similarly to the SUSY searches, also other attempts to look for new physics are so numerous by now that a comprehensive
summary is basically impossible. Many new LHC results have been presented \cite{Sonnenschein}, which show that
exclusion limits for heavy objects, such as heavy vector bosons ($Z',W'$) or excited quarks, have reached the few-TeV range.
Even higher scales are excluded in the context of certain large  extra dimension models or the searches for miniature black holes.
Typical exclusion limits for heavy fermions, such as 4th generation partners, are around half a TeV. A number of spectacular events, 
discovered by such analyses, have been shown at the conference (cf.\ Fig.\ \ref{fig:BSM}, right). For sure, the philosophy
of not leaving any stone unturned, will be pursued at the new 8 TeV LHC run, where the higher centre-of-mass energy leads
to a significant increase of effective luminosity, in particular when searching for very heavy objects.

%%%%
%%%%%%%%%%%%%%%%%%%%%%%%%%%% Higgs
%%%%
\section{Searches for the Higgs Boson}
 \label{sec:Higgs}

\begin{figure}[htb]
\centering
\begin{tabular}{lr}
\includegraphics[width=7cm]{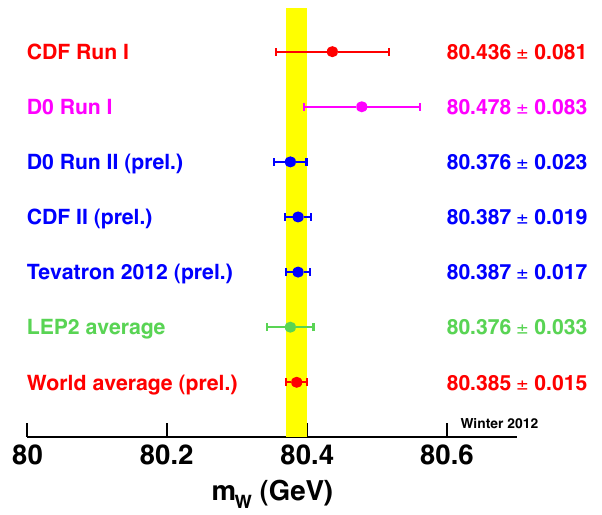} &
\includegraphics[width=7cm]{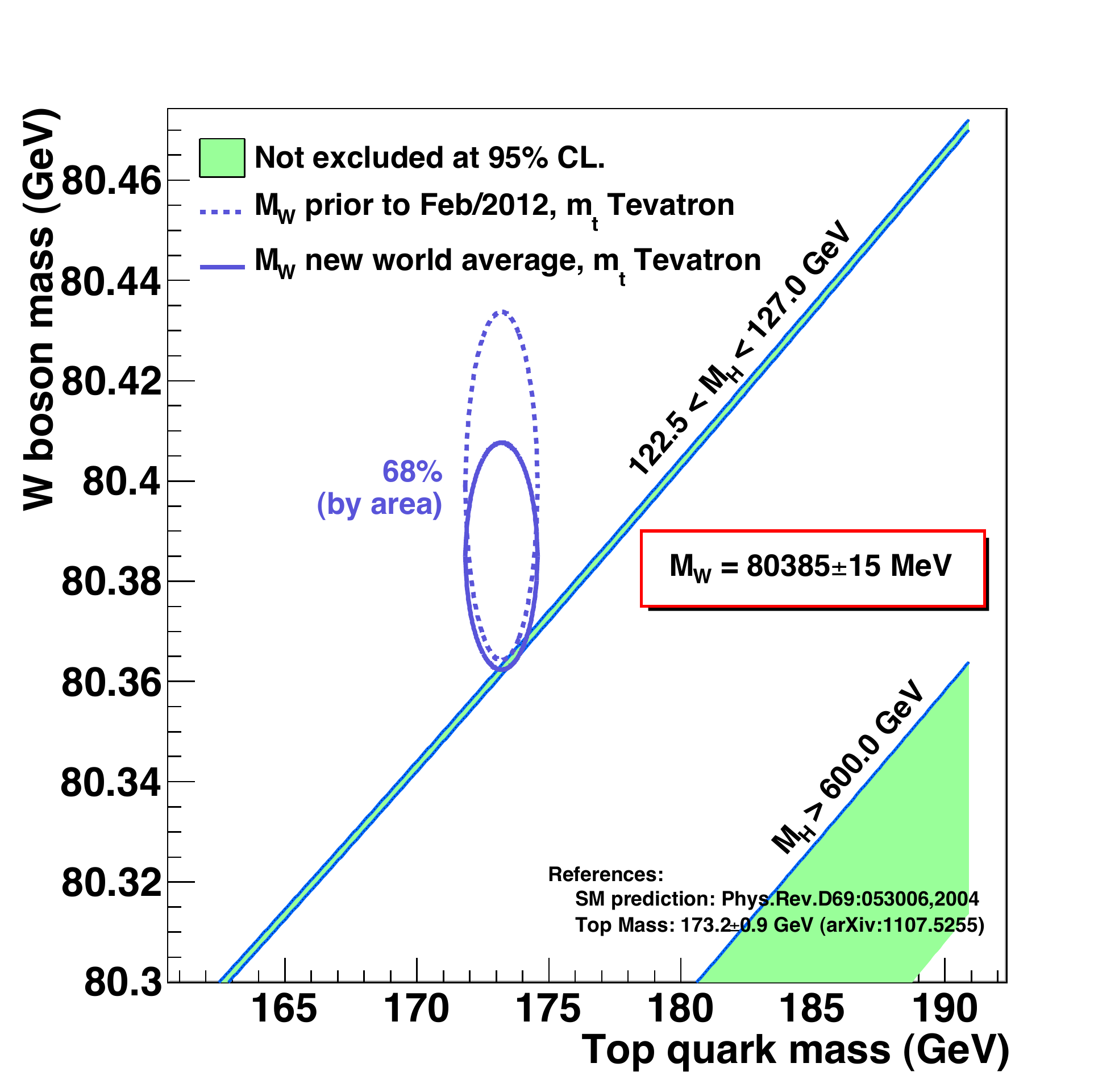} 
\end{tabular}
\caption{Left: Summary of recent measurements of the $W$ mass;
             Right: Consistency check among the $m_W$, $m_t$ measurements, the limits on the Higgs mass,
               and their relation in the context of the SM.
            }
\label{fig:Mw}
\end{figure}

 A traditional approach to testing the electroweak sector of the SM is by looking 
 at the overall consistency among direct measurements of the $W$ and top quark masses,
 current limits on the Higgs mass $m_H$, and the SM relationship among $m_W, m_t$ and $m_H$.
The latest version of this test has been shown at this conference, Fig.\ \ref{fig:Mw}, and can be
considered as one of the real highlights. Indeed, we see that there is consistency, at the 1 sigma level,
among these mass measurements and a possible existence of a SM Higgs with mass around 125 GeV.
The two most important new ingredients to this test are an improved
measurement of $m_W$ at the TEVATRON and the strong Higgs exclusion limits, as discussed below.
The latest, and the world's most precise, determination of the $W$ mass \cite{deSa} has been obtained
by CDF, with an astonishing total uncertainty of 19 MeV, leading to an uncertainty on the latest
TEVATRON combination (world average) of 17 MeV (15 MeV). An important contribution to the uncertainty of 
the final TEVATRON result, related to the knowledge of parton distribution functions, was estimated to be about
10 MeV. However, this error and its possible further reduction in the future, has been questioned by some
of the theorists present \cite{Soper}.
 
\begin{figure}[htb]
\centering
\begin{tabular}{lr}
\includegraphics[width=6.7cm]{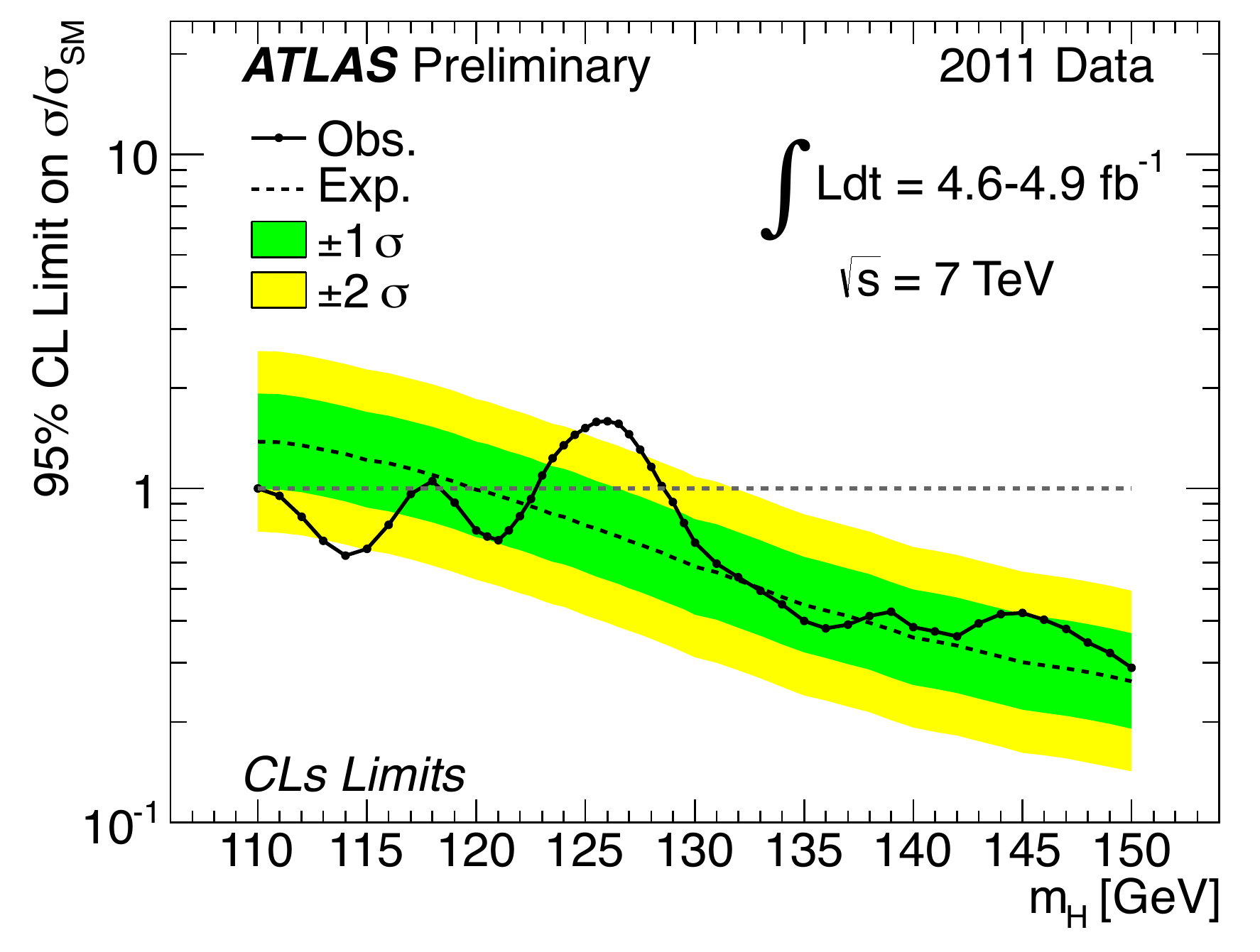} &
\includegraphics[width=7.5cm]{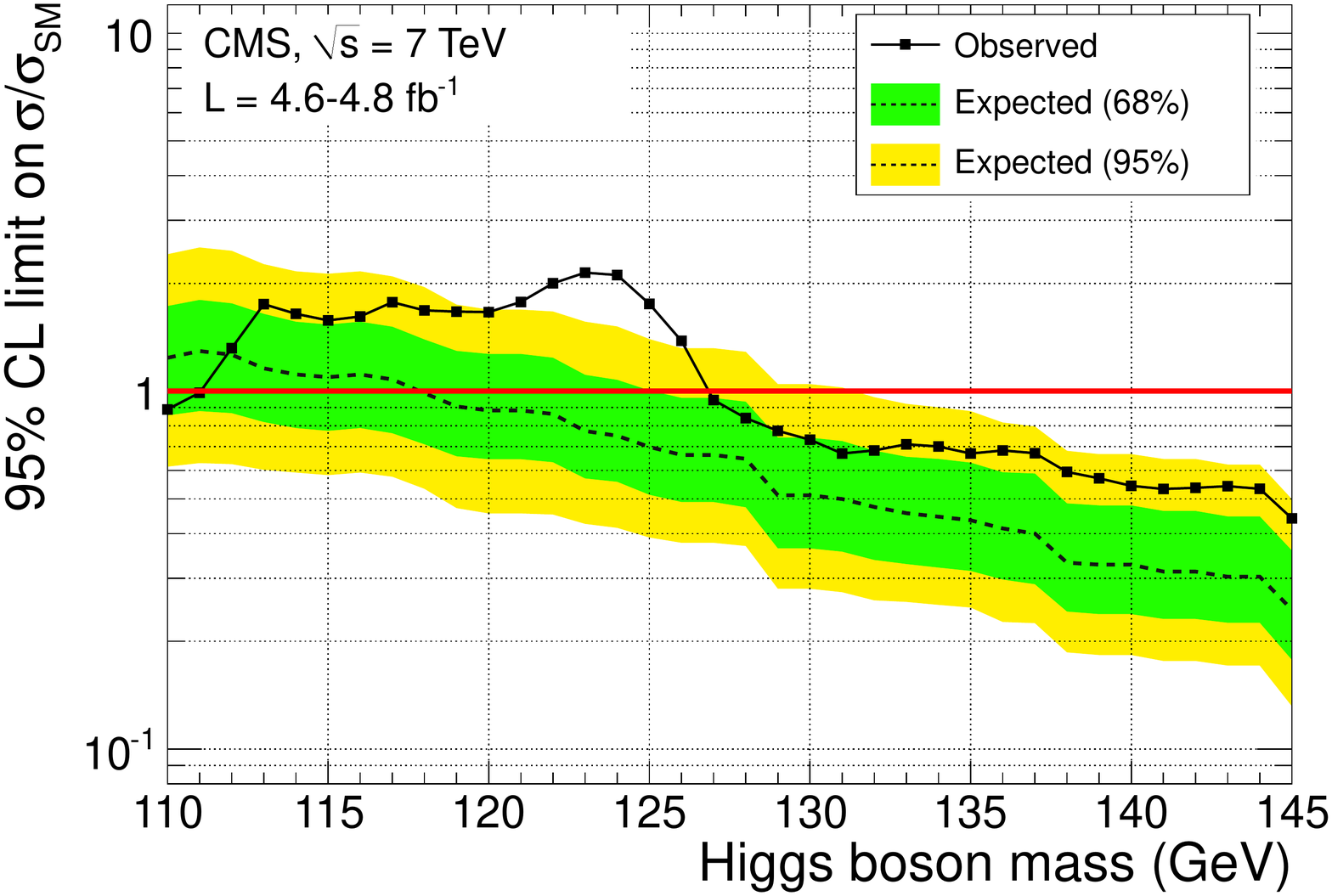} 
\end{tabular}
\caption{95\% C.L.\ exclusion limits on the SM Higgs cross section, as a function of the hypothetical
               Higgs mass, as derived from a combination of the ATLAS (left) and CMS (right) Higgs searches.
}
\label{fig:Higgs_comb}
\end{figure}

\begin{figure}[htb]
\centering
\begin{tabular}{lr}
\includegraphics[width=7cm]{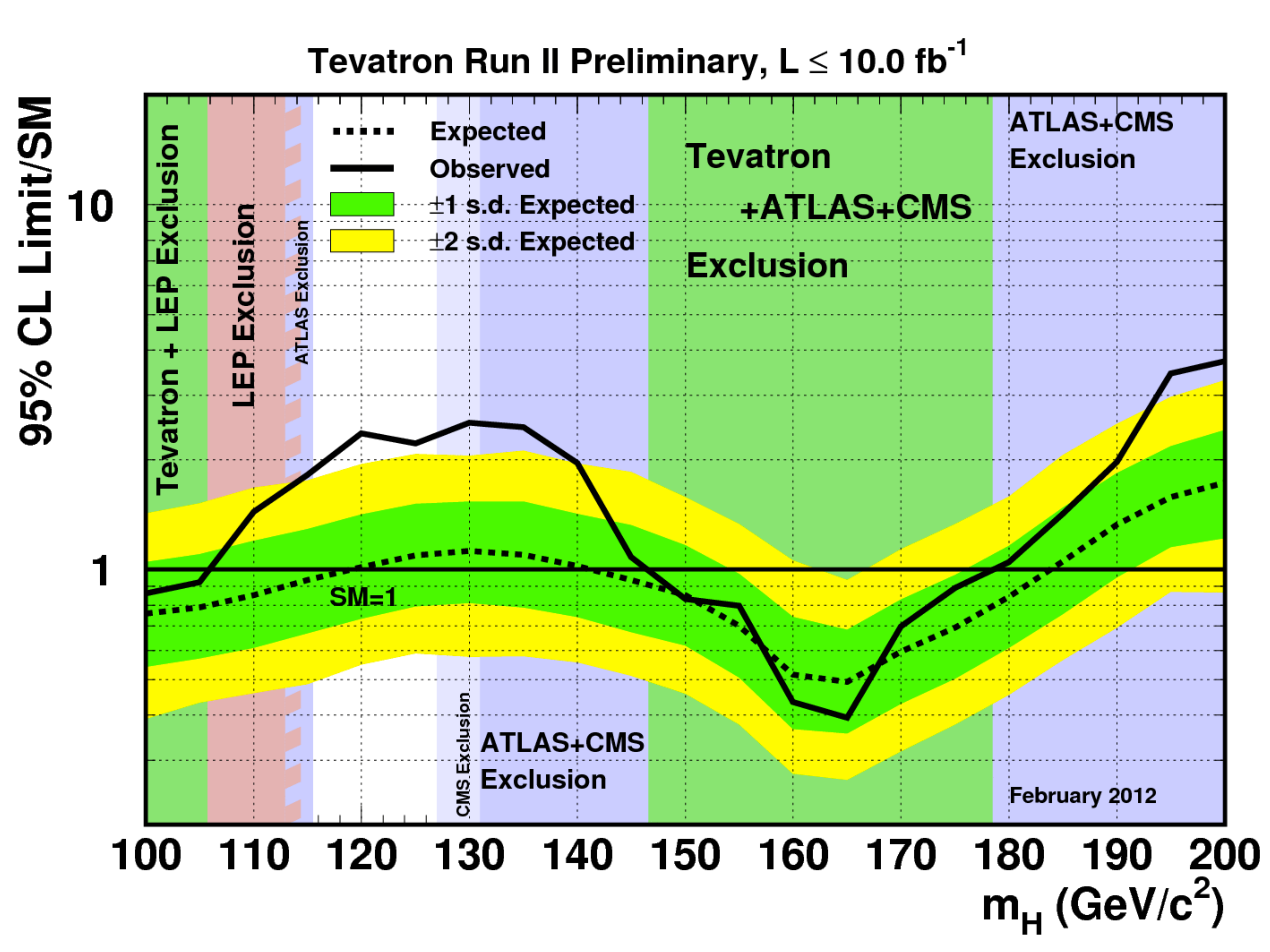} &
\includegraphics[width=7cm]{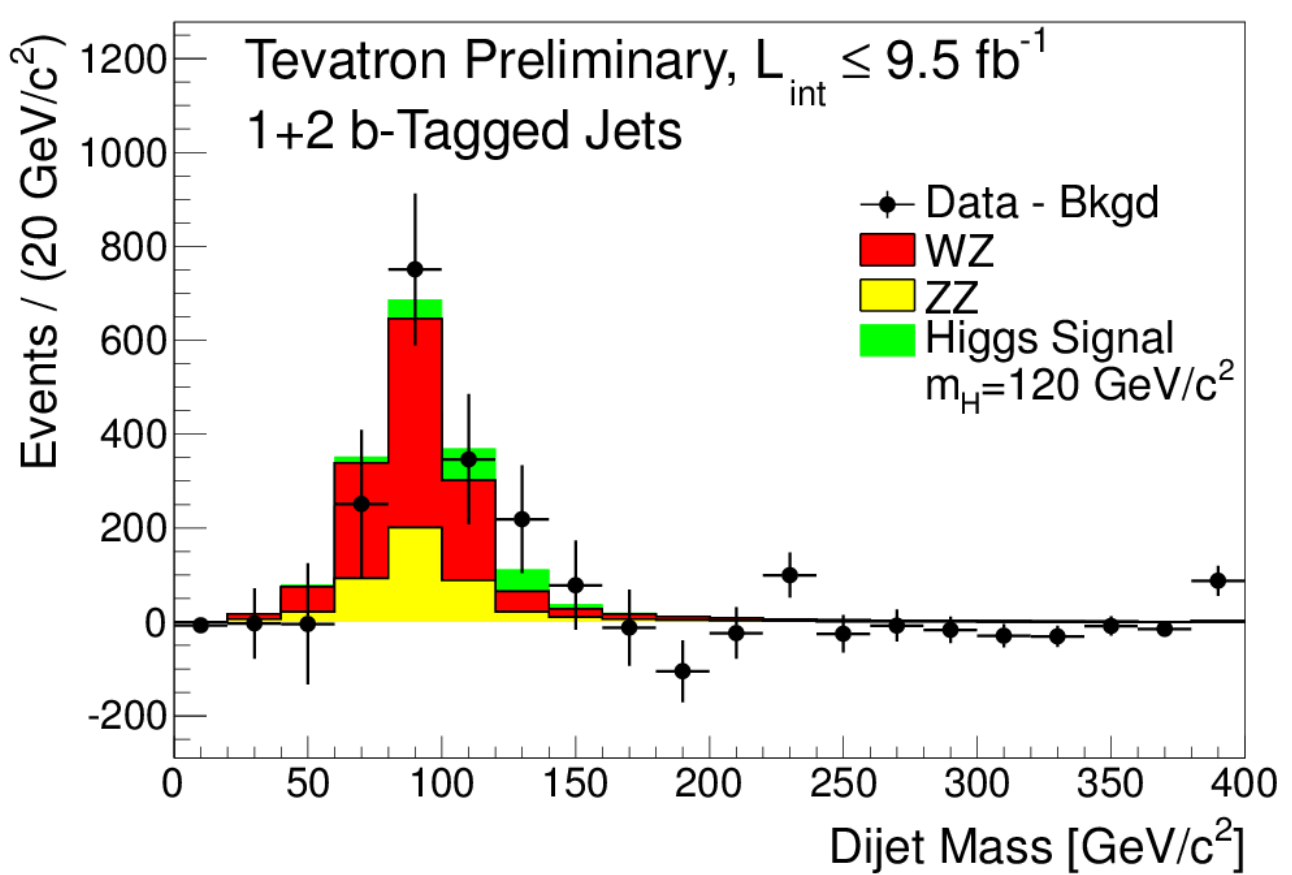} 
\end{tabular}
\caption{Left: 95\% C.L.\ exclusion limits on the SM Higgs cross section, as a function of the hypothetical
               Higgs mass, as derived from a combination of the TEVATRON Higgs searches;              ;
                Right: TEVATRON combination of the measurement of the cross section for $WZ(\rightarrow b\bar b)$ production.}
\label{fig:Higgs_TEV}
\end{figure}

\begin{figure}[htb]
\centering
\begin{tabular}{lr}
\includegraphics[width=7.5cm]{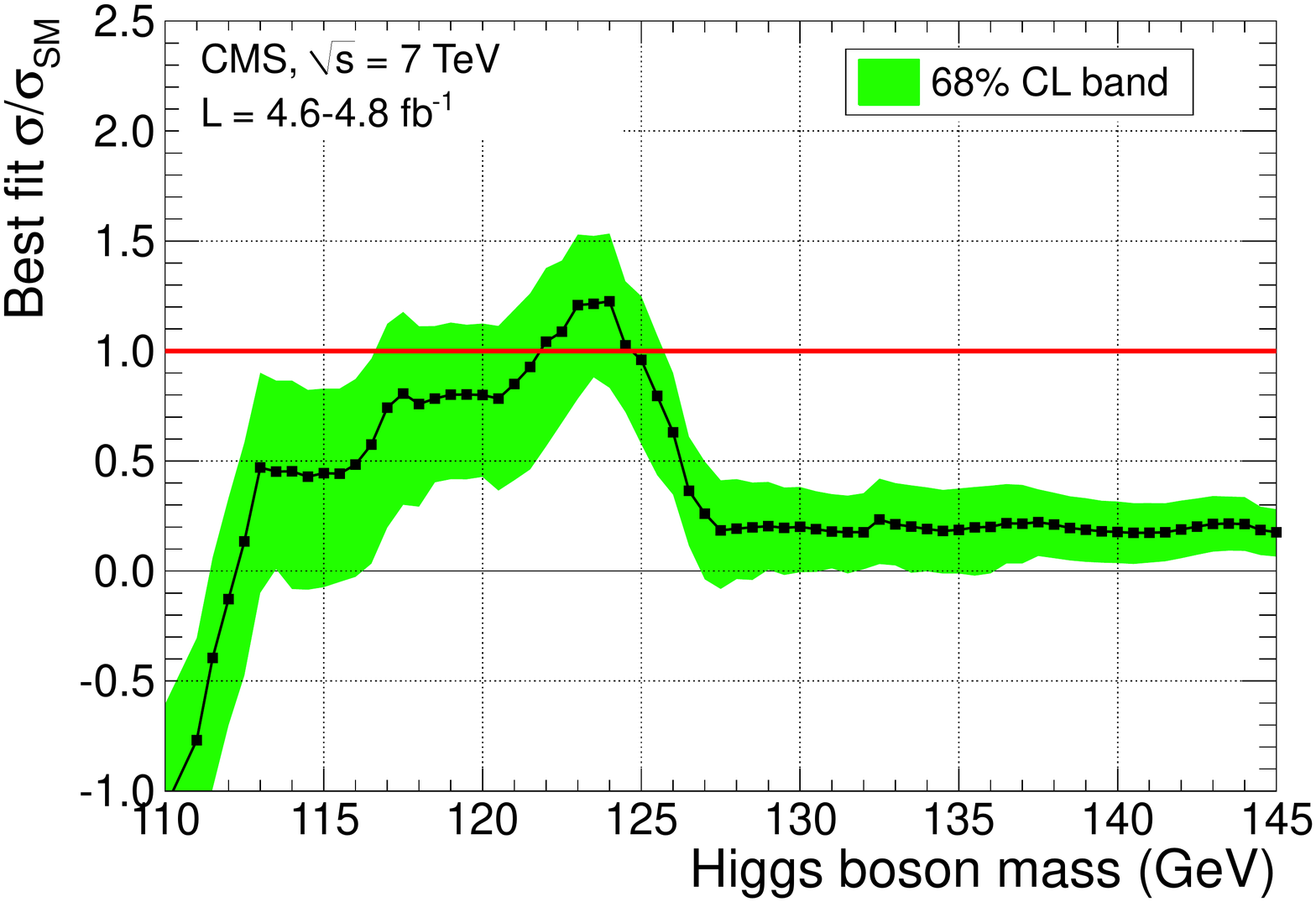} &
\includegraphics[width=6cm]{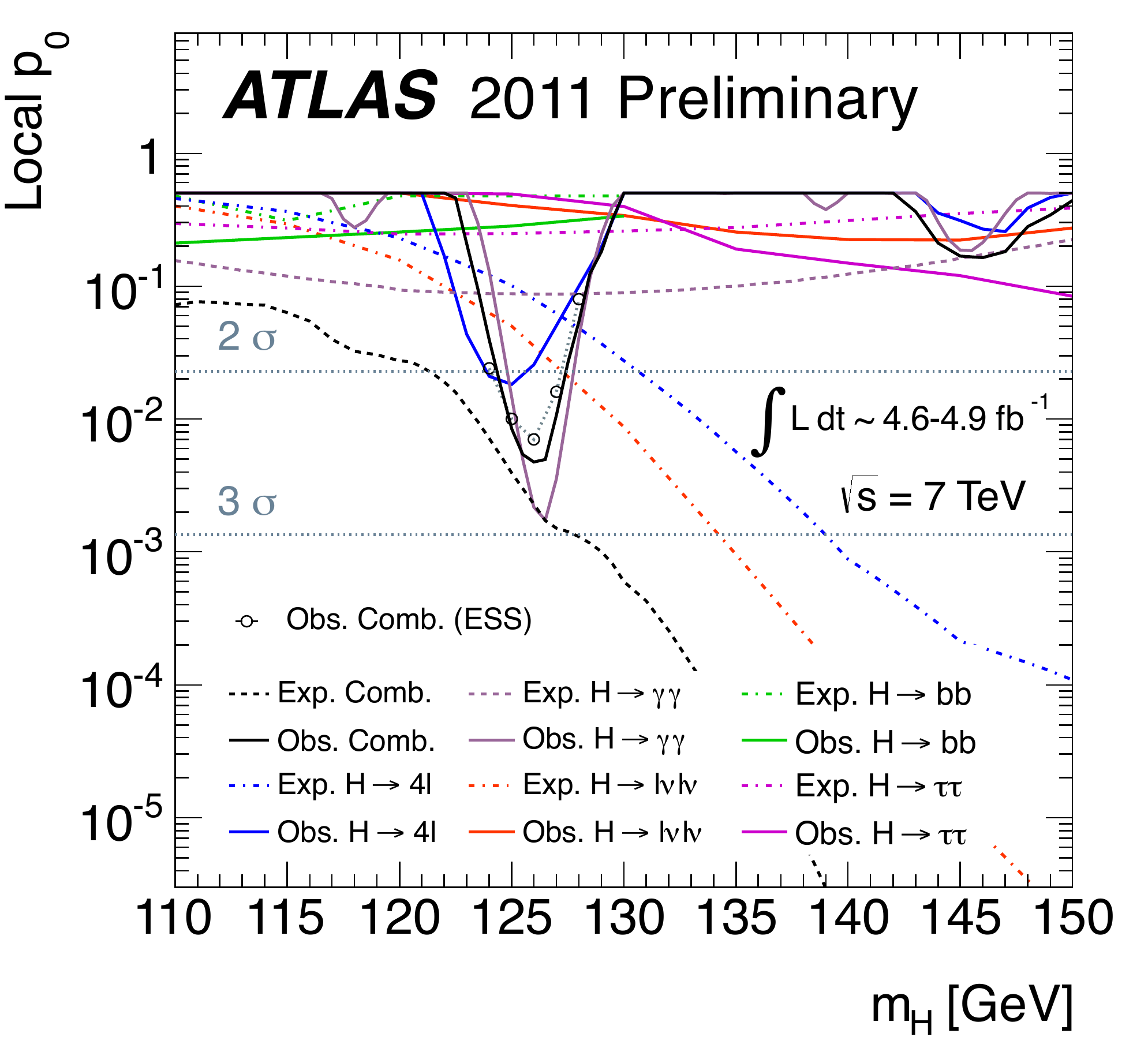} 
\end{tabular}
\caption{Left: Best fit of the signal strength modifier (relative to the SM Higgs cross section),
               as obtained from a combination of the CMS Higgs searches; 
              Right: ATLAS results on the background-only probability for different Higgs search channels.
	}
\label{fig:Higgs_detail}
\end{figure}

The LHC and TEVATRON experiments have presented the latest combinations
of their Higgs searches \cite{Bernhard,Bornheim,Bortoletto}, leading to the following executive summary:
(i) ATLAS excludes, at 95\% C.L., the mass ranges 110-117.5, 118.5-122.5 and 129-539 GeV, (ii) CMS
excludes the range 127.5-600 GeV, and (iii) the TEVATRON has a 95\% exclusion limit for 
$100 < m_H < 106$ and $147 < m_H < 179$ GeV. Very interestingly, all these combined results indicate 
a slight excess in the mass range of roughly 122-128 GeV, with the individual significances of those excesses
somewhat above the 2 sigma level (cf.\ Figs.\ \ref{fig:Higgs_comb} and \ref{fig:Higgs_TEV}). When looking more
closely at the updates presented at this conference, the following observations can be made:
\begin{itemize}
  \item at the CERN seminar on Dec.\ 13, 2011, CMS had presented already a complete set of Higgs searches
          in the various channels, based on the full 2011 statistics. In the meantime, they have performed a new, alternative
           analysis of the $H\rightarrow\gamma\gamma$ channel \cite{Bornheim}, now based on an event classification derived from
             a multi-variate approach to the measurement of photon properties. This leads to a $\sim20\%$ improvement in the
           expected limit, leaving the overall conclusions from the observation on real data unchanged, compared to their earlier analysis.
            Furthermore, they have presented first results for the $WH\rightarrow WWW\rightarrow 3\ell 3\nu$ channel
              and for two new channels in the $H\rightarrow \tau\tau$ search.
  \item in contrast to CMS, previously ATLAS had shown full-2011 statistics results only for the $H\rightarrow\gamma\gamma$ and
           $H\rightarrow ZZ\rightarrow 4\ell$ channels, which are most sensitive in the low-$m_H$ region and characterized by their
              excellent mass resolution. Now, at this conference \cite{Bernhard}, ATLAS has complemented those analyses with a full suite of 
              analyses based on the full 2011 data sample, covering basically all relevant channels and mass regions;
   \item interestingly,  when comparing all those ATLAS and CMS analyses, it is evident that currently both experiments have very similar
             sensitivity in basically all the channels.
   \item the most important recent changes at the TEVATRON \cite{Bortoletto} come from the $VH(\rightarrow b\bar b)$ channel, in particular
             thanks to a considerable improvement in the CDF $b$-tagging performance. 
             An interesting application of the improved tools, and at the same time an important 
              "calibration" channel for the low-mass Higgs search, is their latest measurement of $WZ(\rightarrow b\bar b)$ production, Fig.\ \ref{fig:Higgs_TEV},
                showing excellent agreement with the SM prediction. When analyzing further the recently observed excess in the TEVATRON data,
                one finds that this excess is driven by the $H\rightarrow b\bar b$ search in CDF and by a contribution from D\O\ in the $H\rightarrow WW$ channel.
                The  most significant channel for the CDF excess is $Z(\rightarrow e^+e^-)H(\rightarrow b\bar b)$.
%                where a handful of events with
%                  very high signal over background ratio have been found in the latest 2 fb$^{-1}$ of data registered before the shutdown.
    \item   a closer look at the LHC results reveals that there are downward fluctuations in the measurement of the signal strength modifier 
              (Fig.\ \ref{fig:Higgs_detail}, left)  at the lower  end of the  search region, which should probably looked at with the same attention
                 as the upward fluctuations around 125 GeV, since they tell us something about the (relatively little) statistics still involved. 
                 Furthermore, it was noted that the updated ATLAS result for the $H\rightarrow WW$ channel does not really confirm the excess
                   in the di-photon and four-lepton channel, as for example seen in the distribution of the background-only 
                   probability, Fig.\ \ref{fig:Higgs_detail}, right. However, obviously all these observations correspond to $\sim2$ sigma effects only, thus should
                     be taken with the appropriate grain of salt.                
\end{itemize}

In conclusion, it is simply impressive to see what the LHC and TEVATRON experiments have delivered, in terms of Higgs results,
over such a short time scale between the end of data taking in 2011 and the winter conferences in 2012. A rather solid conclusion appears to
be that a SM Higgs boson is excluded, to very high level of confidence, for masses above $\sim130$ GeV up to about 600 GeV, where the
current searches stop. As mentioned above, all experiments observe some excess in the region around 125 GeV, which may be called
{\em{tantalizing}} at this stage. However, we should not forget the still limited statistics available, correspondingly the still not overwhelming 
significance of these observations, and therefore try not to be carried away. Luckily, new data at 8 TeV start pouring in, the hope being that
  the increased statistics expected in 2012 will allow to make concluding statements on the existence (or exclusion) of a SM Higgs boson. The  challenge is with the analyzers, who will have to avoid, at all costs, the (psychological) bias, which undeniably exists
   after having seen
  the 2011 results. Finally, it is also worth mentioning that a number of non-SM Higgs searches have been 
  presented \cite{Spagnolo,Kroseberg,Couderc}, eg.\ in the context
    of fermiophobic, (N)MSSM or doubly-charged Higgs scenarios, without any significant hints for a signal.

%%%%
%%%%%%%%%%%%%%%%%%%%%%%%%%%% conclusions
%%%%
\section{Conclusions}
 \label{sec:conclusion}

Following the 93 \mbox{(!)} presentations of this conference has been most interesting and allowed
obtaining an excellent and rather complete overview of the present theoretical and experimental
status in the field of QCD and high energy interactions. The wealth of new data, in particular arriving
from the LHC experiments, is overwhelming and exciting at the same time. So far, the Standard Model
appears to be as healthy as ever, with no really significant indication for a deviation from its
predictions observed, and with the final missing building block, the Higgs boson, probably on the horizon.
In a year from now, our big puzzle called "particle physics up to the TeV scale" will be even more 
complete than already seen this year, and we might know then if there is any space left for some
missing piece of the puzzle, entitled "new physics".

%%%%
%%%%%%%%%%%%%%%%%%%%%%%%%%%% thanks
%%%%
\section*{Acknowledgments}
My sincere thanks go to the organizers of this fantastic conference, for their kind invitation
to give this experimental summary talk. It was a real honor for me. The organizers not
only managed, as every year, to run a smooth and very high-quality conference, but also provided sun shine
throughout the whole week (which unfortunately could not be exploited to the maximum 
by the summary speakers). I would also like to thank all the speakers and colleagues,
who provided input for this summary and who answered my many questions.  Finally, very special thanks
go to B.\ Klima and D.\ Treille for their comments on the manuscript.

%%%%
%%%%%%%%%%%%%%%%%%%%%%%%%%%% bibliography
%%%%

\end{document}